\documentclass[fleqn,usenatbib]{mnras}

\usepackage{newtxtext,newtxmath}

\usepackage[T1]{fontenc}

\DeclareRobustCommand{\VAN}[3]{#2}
\let\VANthebibliography\thebibliography
\def\thebibliography{\DeclareRobustCommand{\VAN}[3]{##3}\VANthebibliography}

\usepackage{graphicx}	
\usepackage{amsmath}	

\usepackage{soul} 

\defcitealias{K5_Bekki2013}{B13}
\defcitealias{K8_Tsujimoto_et_al1995}{T95}
\defcitealias{K9_Van_den_hoek_Groenewegen1997}{VG97}

\title[Simulations of Massive Star Clusters]{Star cluster formation from giant molecular clouds in the Small Magellanic Cloud about 2 Gyr ago: their origin, structures, and kinematics}

\author[M. Williams, K. Bekki \& M. McKenzie]{
Mia L. Williams,$^{1}$\thanks{E-mail: 22711863@student.uwa.edu.au}
Kenji Bekki,$^{1}$
Madeleine McKenzie,$^{1,2}$
\\
$^{1}$ICRAR M468, The University of Western Australia, 35 Stirling Hwy, Crawley, WA 6009, Australia\\
$^{2}$Research School of Astronomy and Astrophysics, Australian National University, Canberra, ACT 2611, Australia\\
}

\date{Accepted XXX. Received YYY; in original form ZZZ}

\pubyear{XXX}

\begin{document}
\label{firstpage}
\pagerange{\pageref{firstpage}--\pageref{lastpage}}
\maketitle

\begin{abstract}
    Recent observations have found that the age distribution of star clusters (SCs) in the Small Magellanic Cloud (SMC) shows a sharp peak around 2 Gyr ago. However, it is theoretically unclear what physical processes are responsible for such sudden formation of SCs in the SMC. Here we investigate whether massive SCs with initial masses more than $10^5$ $\text{M}_\odot$ can be formed during tidal interaction of the SMC with the Large Magellanic Cloud (LMC) about 2 Gyr ago, based on our new simulations, which include molecular hydrogen formation on dust grains and SC formation within giant molecular clouds (GMCs). We find that the formation of GMCs with masses more than $10^5$ $\text{M}_\odot$ can be dramatically enhanced due to the tidal force of the LMC-SMC interaction. We also find that gravitationally bound massive SCs can be formed within these GMCs, though their mean stellar densities ($10^4$ $\text{M}_\odot \text{pc}^{-3}$) are systematically lower than those of the genuine globular clusters (GCs). All simulated SCs have diffuse extended stellar envelopes that were formed from multiple merging of sub clusters within their natal GMCs. Furthermore, we find that some of the simulated SCs can have considerable global internal rotation and substructures surrounding them. Based on these simulation results, we discuss the origin of the observed diverse properties of SCs in the SMC and the physical roles of galaxy interaction in the formation of massive SCs from GMCs.
\end{abstract}




\section{Introduction}
The Large and Small Magellanic Clouds, due to their proximity to our Galaxy, offer a unique opportunity to examine the roles of their tidal interactions in the formation and evolution of various stellar and gaseous objects (e.g.,\citealt{K11_Westerlund1997}; \citealt{K43_DOnghia_Fox2016}). The two irregular dwarf galaxies are interacting with the Galaxy, which could have influenced the evolution of these three systems (e.g., \citealt{78Murai_Fujimoto1980}; \citealt{K12_Weinberg2000}, \citealt{69Yoshizawa_Noguchi2003}; \citealt{K13_Besla_et_al2012}; \citealt{K3_Diaz_Bekki2012}). 
Two of the most recent LMC-SMC interactions that possibly occurred ${\sim}0.2$, and ${\sim}2$ Gyr ago (e.g., \citealt{69Yoshizawa_Noguchi2003}; \citealt{20Harris_Zaritsky2004}; \citealt{1Rafelski_Zaritsky2005}; \citealt{29Bekki_Chiba2007}; \citealt{17Bekki_Chiba2009}; \citealt{11Bitsakis_et_al2018}; \citealt{39Nayak_et_al2018}) have been investigated by astronomers in the contexts of the formation of the Magellanic Stream (\citealt{K15_Gardiner_Noguchi1996}; \citealt{67Sawa_Fujimoto_Kumai1999}; \citealt{89Pardy_DOnghia_Fox2018}), its leading arms (\citealt{M6_Cullinane_et_al2021}) and the Bridge (\citealt{24Zivick_et_al2019}), as well as the chemical and dynamical evolution of the Clouds (\citealt{M8_Bekki_Chiba2005}; \citealt{M9_Patel_et_al2020}).

The LMC and SMC are ideal for studying the origin of star clusters (SCs), because their various physical properties such as size, dynamical mass, stellar kinematics, stellar populations, age and metallicity have been observed by many previous works (e.g., \citealt{27Grebel_et_al1999}; \citealt{5Da_Costa2002}; \citealt{40Glatt_Grebel_Koch2010}; \citealt{56Perren_Piatti_V2017}; \citealt{53Piatti2020}). Despite these works on SCs, there are a number of key problems yet to be resolved. For example, SCs with ages from 4-12 Gyrs in the LMC ("age gap"; \citealt{K14_Da_Costa1991}) have minimal observations (only ESO 121-SC03), and the origin is yet to be fully resolved, though numerical simulations of the LMC suggested that reactivation of SC formation due to LMC-SMC interaction could be the origin. The SCs of the SMC are useful for studying how tidal interaction of galaxies can influence the formation and evolution of SCs, given that the SMC has had clear interactions with the LMC. There are many previous studies which look specifically at SMC SCs.

For example, in \cite{11Bitsakis_et_al2018}, there are $1319$ SCs surveyed in the central and outer regions of the SMC. Their catalog shows the SCs to have an average radius of ${\sim}16$ pc, with values ranging from $6$ pc to $30$ pc. From further analysis of the SC's ages, peaks of cluster formation at $0.04$, $0.27$ and $2$ Gyr are detected for outskirts of the SMC, and peaks of $0.02$, $0.2$ and $0.8$ Gyr ago in the bar region. Their peak periods also demonstrate the tendency for younger SCs to be in the central bar region, and older SCs to be in the halo (\citealt{1Rafelski_Zaritsky2005}; \citealt{39Nayak_et_al2018}). In comparison, \cite{64Piatti_et_al2008} looks at $7$ SCs in the SMC's central region, and notes a large number of clusters aged between $2.5$ and $1$ Gyr, further supported by \cite{1Rafelski_Zaritsky2005}, who found with their survey of 195 SCs in the SMC the same peak at ${\sim}2$ Gyr ago occurred.

Despite the correlation between enhanced SC formation periods and the LMC-SMC interactions being noted
(e.g., \citealt{27Grebel_et_al1999}; \citealt{3Parmentier_de_Grijs_Gilmore2003}; \citealt{1Rafelski_Zaritsky2005}; \citealt{16Chiosi_et_al2006}; \citealt{40Glatt_Grebel_Koch2010}; \citealt{37Maji_et_al2017}; \citealt{11Bitsakis_et_al2018}; \citealt{39Nayak_et_al2018}), only a few theoretical studies on SC formation in the Magellanic Clouds (MCs) have been done. For example, \cite{78Murai_Fujimoto1980} suggested that interactions between the LMC and SMC could result in bursts of star formation. Through analysis of the MCs amongst other galaxies, \cite{73Kumai_Basu_Fujimoto1993} found most galaxies with Globular Clusters (GCs) less than $0.1$ Gyr old (young GCs) have strong interactions in their history. Formation of single or binary SCs due to high-speed cloud-cloud collisions triggered by LMC-SMC interaction was numerically investigated in detail by \cite{79Bekki_et_al2004}. However, these previous studies did not resolve well the formation of SCs from GMCs in their galaxy-scale models and simulations. It is thus unclear how gravitationally bound SCs consisting of many individual stars can really form in the interacting Clouds.

\begin{table}
\centering
\caption{Description of the basic parameter values
for the fiducial tidal interaction model, T1. (\citealt{K8_Tsujimoto_et_al1995}; \citetalias{K8_Tsujimoto_et_al1995}, \citealt{K9_Van_den_hoek_Groenewegen1997}; \citetalias{K9_Van_den_hoek_Groenewegen1997}, \citealt{K5_Bekki2013}; \citetalias{K5_Bekki2013})}. 

\label{table:1KParameters}
\begin{tabular}{ll}
\hline
{Physical properties}
& {Parameter values}\\
\hline
Dark matter structures (LMC and SMC)
& NFW  \\
Total halo mass (LMC)
& $6.0 \times 10^{10} {\rm M}_{\odot}$  \\
Galaxy virial radius (LMC)
&  60.1 kpc  \\
$c$ parameter of galaxy halo
&  $c=12$  \\
Stellar disk  mass (LMC) & $3.0 \times 10^{9} {\rm M}_{\odot}$     \\
Disk scale height (LMC) & $0.3$ kpc \\
Disk scale length (LMC) & $3.5$ kpc \\
Total halo mass (SMC)
& $4.8 \times 10^{9} {\rm M}_{\odot}$  \\
Stellar disk  mass (SMC) & $6 \times 10^{8} {\rm M}_{\odot}$     \\
Stellar disk size (SMC) & $2.63$ kpc \\
Gas disk size (SMC) & $5.26$ kpc \\
Disk scale length (SMC) & $0.35$ kpc \\
Gas fraction in a disk & 0.5     \\
Mass resolution & $3.0 \times 10^3 {\rm M}_{\odot}$ \\
Size resolution & 18.9 pc \\
Star formation law & The KS law \\
& with $\rho_{\rm th}= 100$ ${\rm cm}^{-3}$  \\
Initial central metallicity   &   ${\rm [Fe/H]_0}=-0.6$ \\
Initial  metallicity gradient   &   $\alpha_{\rm d}=-0.01$ dex kpc$^{-1}$ \\
Chemical yield  &  \citetalias{K8_Tsujimoto_et_al1995} for SN,\\
&\citetalias{K9_Van_den_hoek_Groenewegen1997} for AGB \\
Dust yield  &  \citetalias{K5_Bekki2013}\\
Dust formation model
& $\tau_{\rm acc}=0.25$ Gyr,   \\
 & $\tau_{\rm dest}=0.5$ Gyr  \\
Initial dust/metal ratio  & 0.4  \\
${\rm H_2}$ formation
& Dependent on dust \\
 & abundance (B13) \\
Feedback
& SNIa and SNII  (no AGN) \\
\hline
\end{tabular}
\end{table}

The question of why and how such a large number of SCs were suddenly formed in the SMC ${\sim}2$ Gyr ago is of particular importance theoretically, because 2 Gyr ago might be a major epoch of LMC-SMC interaction (e.g., \citealt{K15_Gardiner_Noguchi1996}; \citealt{69Yoshizawa_Noguchi2003}). Also, 2 Gyr ago might be the period when the SMC started to be influenced by warm halo gas of the Galaxy, which might have further influenced the evolution of GMCs of the SMC: ram pressure of the Galactic halo gas can trigger the enhanced star formation in the LMC gas disk (\citealt{K16_Mastropietro_Burkert_Moore2009}). Firstly, we need to understand the physical mechanisms by which the formation of massive SCs became possible in the SMC ${\sim}2$ Gyr ago, and secondly, why 2 Gyr ago is the particular epoch for the SC formation in the SMC based on realistic numerical simulations that include SC formation from GMCs.

Internal rotation and velocity dispersion of SCs in the LMC and SMC can provide useful information on the formation processes of SCs, however, there are only a few observational studies which investigated the internal stellar kinematics. For example, \cite{K18_Mackey_et_al2013} investigated the line of sight velocities of 21 stars in the LMC SC NGC 1846 with its extended main-sequence turnoff, which is a sign of multiple stellar populations (e.g., \citealt{K19_Keller_Mackey_Da_Costa2011}), and thereby found a clear sign of internal rotation in the cluster. Although such rotational dynamics have been discussed by recent observational works on the Galactic GCs (e.g., \citealt{K20_Bianchini_et_al2018}; \citealt{K21_Barth_et_al2020}), the origin of internal rotation in the SCs of the LMC and the SMC has not been discussed: it is not observationally clear whether SCs of the SMC have such internal rotation. Therefore, it would be important for the theory of SC formation in the SMC to investigate whether the simulated SCs can have such rotation, and in what physical conditions the SC can have rotation.

We have already investigated a number of key physical processes related
to SC formation such as GMC collisions during the LMC-SMC collision
(e.g., \citealt{79Bekki_et_al2004}), gas accretion onto SCs (e.g., \citealt{K32_Bekki_Mackey2009}; \citealt{K33_Armstrong_et_al2018}), new star formation within massive globular clusters accreting gas from ISM  (e.g., \citealt{K34_McKenzie_Bekki2018}; \citealt{K23_McKenzie_Bekki2021}), and the roles of gas infall from the SMC to the LMC in SC formation (\citealt{K35_Bekki_Chiba2007}; \citealt{K36_Tsuge_et_al2019}). We have not yet investigated the entire formation processes of SC formation from their natal GMCs.

In this paper, we aim to investigate both the formation of GMCs in the gaseous disk of the SMC and the subsequent SC formation within the GMCs in a self-consistent way, using our new original numerical simulations of the LMC-SMC tidal interaction. These new simulations include the formation of molecular hydrogen on dust grains (\citealt{K5_Bekki2013}; \citealt{K6_Bekki2015}) so that they can predict the formation processes of GMCs consisting largely of molecular gas. Furthermore, we introduce a new two-stage SC formation model in which the formation of a SC with many ($>10^4$) individual stellar particles (with a mass resolution of the order of 1 ${\rm M}_{\odot}$) in a GMC can be investigated in detail. In the present study, we assume that the LMC-SMC interaction happened ${\sim} 2$ Gyr ago and thereby investigate how LMC-SMC tidal interactions can enhance the formation of GMCs, and thus the formation of the massive SCs with masses more than $10^5$ $\text{M}_\odot$ in the SMC.

\begin{figure*}
 \centering
 \includegraphics[width=15cm]{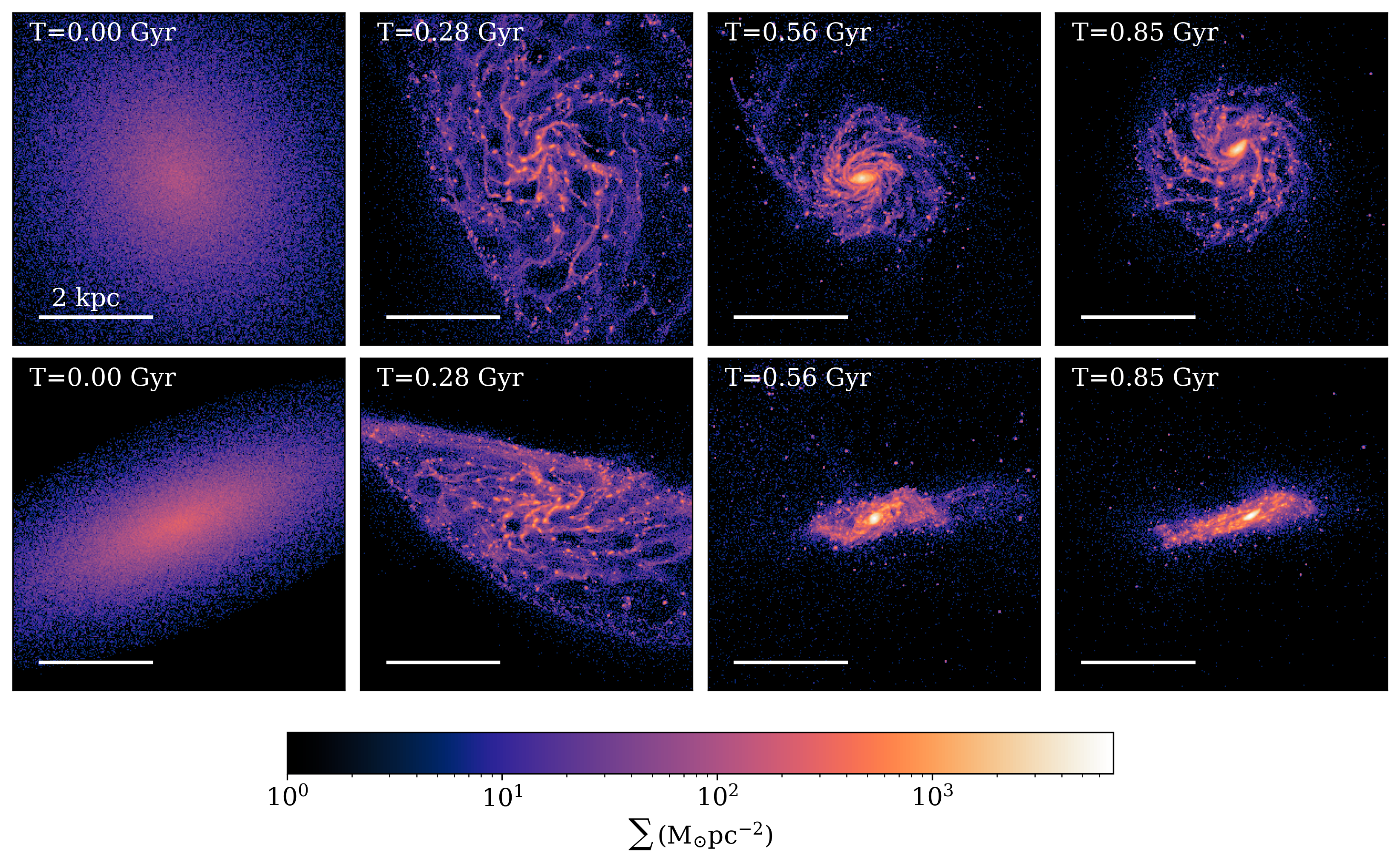}
 \caption{Time evolution of surface gas density in the SMC disk projected onto the x-y plane (upper row) and onto the y-z plane (lower row) in the fiducial model T1.
 The scale bar of 2 kpc is shown in the lower left corner and the time T is shown in the upper right corner. Binning size is 20 pc, and the colour bar shows gas density in units of $\text{M}_\odot \text{pc}^{-2}$.}
 \label{fig:1SMCall}
\end{figure*}

Although the long-term evolution of SCs through stellar encounters (two-body relaxation effects) is quite important for the dynamical evolution of SCs (such as tidal disruption by their host galaxies; \citealt{K39_Lamers_Gieles_Portegies_Zwart2005}) and age/mass distribution of SCs (e.g., \citealt{K42_Baumgardt_et_al2013} for the LMC case), we do not investigate this key issue in the present study. This is mainly because we need to focus on the formation processes of SCs in the SMC triggered by the LMC-SMC interaction. There are several key observational results that we do not address, such as the age or metallicity radial gradients of SCs in the SMC (e.g., \citealt{K37_Parisi_et_al2015}; \citealt{K40_Dias_et_al2016}),  various cluster core radii in the SMC SCs (\citealt{K38_Santos_et_al2020}),  new SC populations in the Magellanic Bridge (e.g., \citealt{K41_Bica_et_al2020}). We will discuss these other observations in our future papers, because discussing these in the same paper would be too complicated and excessive.

The plan of this paper is as follows. In \S \ref{sec:the_model}, we describe in detail the new two-step numerical model for the formation of SCs of the SMC. In \S \ref{sec:results}, we present the effect of tidal interactions on the creation of GMCs, and the physical properties of the simulated SCs where such interactions were present; such
as their internal structures and kinematics. In \S \ref{sec:discussion}, we discuss the physical origins of the 2 Gyr burst SC population of SCs and the observational implications of the present new results. We summarize our new results in \S \ref{sec:conclusion}. 
\section{The Model}
\label{sec:the_model}
\subsection{Two-step investigation}
It is not practical for the present study to investigate the formation of every single SC with the total number of stars ($N$) being ${\sim}10^5$ in the SMC, owing to strong limitations in mass and size resolutions of galaxy-scale simulations. Therefore, we adopt the following two-step investigation in the present study. 
We first try to identify the possible formation sites of massive SCs in the SMC by investigating whether each local region of the SMC (e.g., 50 pc $\times$ 50pc) contains a clump of stars with the total mass more than $10^5-10^6$ ${\rm M}_{\odot}$.
In this first step, the dynamical evolution of individual stars (i.e., with masses of $\sim 1 {\rm M}_{\odot}$) cannot be resolved in
the galaxy-scale simulations. However, the possible total number and locations of massive SCs can be investigated in detail for each model.

Our second step is to investigate just one SC from its formation within a Giant Molecular Cloud to its stripping processes. 
In this second step, the formation of 
a high-density GMC is identified, and the GMC is converted into numerous stars with $N\approx 10^5$ to achieve better resolution. 
Accordingly, the early dynamical evolution of the new SC within their natal GMC can be investigated, though the long-term evolution of the SC through two-body relaxation
effects cannot be investigated owing to the adopted simulation code. The total mass, size, shape, and stellar kinematics of a SC can be derived in this second step so that we can discuss whether the simulated bound SCs can be similar to the observed ones in the SMC for various physical properties.
The selection process of a possible SC-forming GMC in a simulation is described later in this paper.

\subsection{Chemodynamics in the SMC}

We assume that the SMC is a disk galaxy embedded in a massive dark matter halo, though it is observationally unclear whether the SMC contains a significant amount of dark matter within its optical radius of 3 kpc (\citealt{K1_Bekki_Stanimirovic2009}). The SMC is accordingly composed of dark matter halo, stellar disk, and gaseous disk (without a bulge) in the present study.
The total masses of its dark matter, stars, and gas
are denoted as $M_{\rm h}$, $M_{\rm s}$, and $M_{\rm g}$, respectively. We mainly investigate the model with 
the total mass of the SMC ($M_{\rm t}$) being $6 \times 10^9$ $M_{\odot}$, the dark matter fraction of 0.8, and the gas mass fraction ($f_{\rm g}$) of 0.5. As adopted in our previous simulations of the SMC interacting with the LMC
and the Galaxy (\citealt{29Bekki_Chiba2007}; \citealt{K2_Bekki_Chiba2008}; \citealt{K3_Diaz_Bekki2012}),
the SMC's dark matter is truncated around $R=5.25$ kpc, owing to its interaction with the LMC. The lower dark matter fraction allows the SMC disk to form a stellar bar
within several dynamical timescales in the present study.
We also investigate the models with different $M_{\rm t}$ just for comparison.

The `NFW' profile for the dark matter halo (\citealt{K4_Navarro_Frenk_White1996}) suggested from cold dark matter simulations is adopted for all models in the present study. We assume that the stellar and gaseous disks of the SMC are represented by the standard exponential profile, and the radial ($R$) and vertical ($Z$) density profiles of the adopted disk are
assumed to be proportional to $\exp (-R/R_{0}) $ with scale length $R_{0} = 0.2R_{\rm s}$  and to ${\rm sech}^2 (Z/Z_{0})$ with scale length $Z_{0} = 0.04R_{\rm s}$, respectively, where $R_{\rm s}$ is the stellar disk size of the SMC.
The gas disk with a size  $R_{\rm g}=2R_{\rm s}$
has the  radial and vertical scale lengths
of $0.2R_{\rm g}$ and $0.02R_{\rm g}$, respectively.
We investigate the models with $R_{\rm s}=2.63$ kpc,
which is similar to the observed size of the present-day SMC.
The more extended gas disk (${\sim} 5$ kpc) is required to explain the observed spatial distribution of the Magellanic Stream (\citealt{K3_Diaz_Bekki2012}), though it is larger than the present-day gas disk size of the SMC. 
The initial radial and azimuthal velocity dispersion's are assigned to the SMC disk according to the epicyclic theory with Toomre's parameter $Q$ = 1.5.

The present study uses our original simulation code developed by \cite{K5_Bekki2013} and \cite{K6_Bekki2015}, which is a Nbody/SPH simulation code and includes various physical processes of interstellar medium (ISM) in galaxies. 
Therefore, the present simulations include
star formation, chemical evolution, dust evolution, metallicity-dependent radiative cooling, feedback effects of supernovae, and formation of molecular gas in a fully self-consistent manner.
Since the details of these models are given in the above two papers, we will only briefly describe these in the present study.
The Kennicutt-Schmidt law for galaxy-wide star formation
(\citealt{K7_Kennicutt1998}) is adopted with the threshold gas density for star formation being 100 atom cm$^{-3}$. The initial central metallicity of gas in a disk ([Fe/H]$_0$) in the SMC is set to be $-0.6$  and the radial metallicity gradient is $-0.01$ dex kpc$^{-1}$,
which is a bit shallower than the Galactic value ($-0.04$) (\citealt{K22_Choudhury_et_al2018}).

Since the formation of SC-forming GMCs is key in the present simulations, the formation of molecular hydrogen from neutral hydrogen on dust grains is included, using the dust abundance of gas and the interstellar
radiation field around the gas, and estimated for each SPH gas particle. Chemical yields for SNIa and SNII and those for AGB stars are adopted from \cite[T95]{K8_Tsujimoto_et_al1995}  and
\cite[VG97]{K9_Van_den_hoek_Groenewegen1997} respectively.
The dust growth ($\tau_{\rm acc}$) and destruction ($\tau_{\rm dest}$) timescales are set to $\tau_{\rm acc}=0.25$ Gyr and $\tau_{\rm dest}=0.5$ Gyr.
The canonical Salpeter initial mass function (IMF) of stars with the exponent of IMF being $-2.35$ is adopted. These parameter values are all the same as those
adopted in our previous works (\citealt{K5_Bekki2013}). The photoelectric heating effects, which are included in our others studies of luminosity disk galaxies (\citealt{K10_Osman_Bekki_Cortese2020}), are not included in the present study.

\subsection{The fixed gravitational potential of the LMC}

\begin{figure}
 \includegraphics[width=\columnwidth]{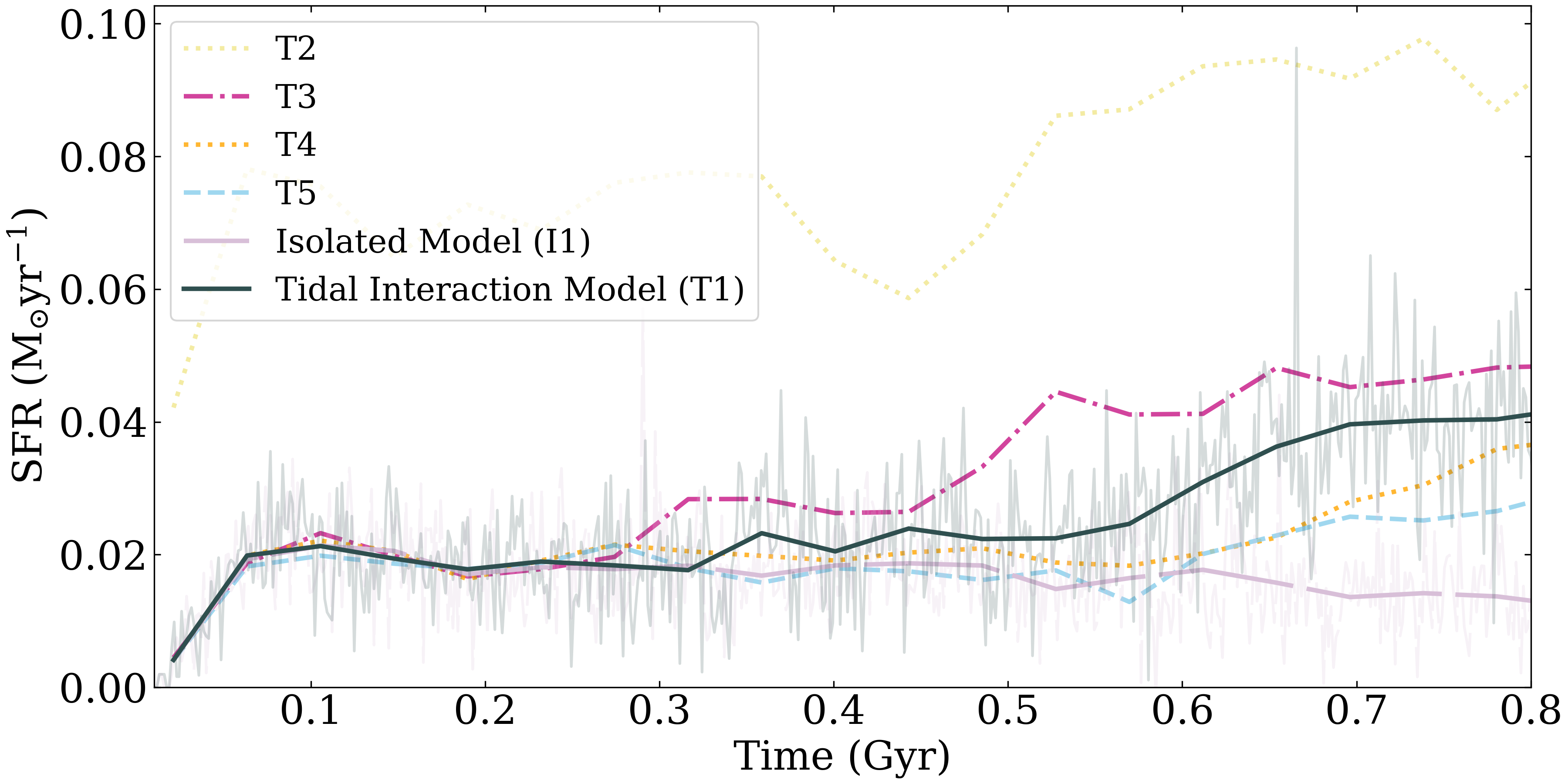}
 \caption{Time evolution of SFR over 0.8 Gyr in the fiducial interacting model T1 (teal solid) and isolated model I1 (purple dashed). 
 Darker lines show average SFR as taken in 20 step intervals, with the same average also shown for the remaining four interacting models (listed in Table \ref{table:1SMCmod}). Lighter lines for T1 and I1 show original SFR values from the simulation.
 Note that, comparing the T1 to I1 models, only the T1 model shows the enhancement of SFR after the LMC-SMC tidal interaction.}
 \label{fig:2SFR}
\end{figure}

The LMC is assumed to influence
the orbit of the SMC through a \emph{fixed} gravitational potential of the two components: a disk, and an extended dark matter halo.  
The dark matter halo is the dominant component of the LMC
and it is modeled by using the NFW density distribution:
\begin{equation}
{\rho}_{\rm LMC} (r)=\frac{\rho_{0}}{(cr/R_{\rm vir})(1+cr/R_{\rm vir})^2},
\end{equation}
where $r$ is the spherical radius, $\rho_{0}$ is the characteristic density, $c$ is the concentration parameter, and $R_{\rm vir}$ is the virial radius. The total mass within $r=R_{\rm vir}$ is called the virial mass and is given by
\begin{equation}
M_{\rm vir}=4\pi \rho_{0} (R_{\rm vir}/c)^3( \ln(1+c) - c/(1+c)).
\end{equation}
The concentration parameter of the NFW profile is assumed to be 12 and 
$R_{\rm vir}$ and $M_{\rm vir}$ are set to be free parameters.
We mainly investigate the LMC models
with 
$R_{\rm vir}=60.1$ kpc and $M_{\rm vir}=6 \times 10^{10}$ ${\rm M}_{\odot}$.

The LMC disk is assumed to be represented by the \cite{K17_Miyamoto_Nagai1975} potential
\begin{equation}
{\Phi}_{\rm disk}=-\frac{GM_{\rm d}}{\sqrt{R^2 +{(a+\sqrt{z^2+b^2})}^2}},
\end{equation}
where $M_{\rm d}$ is the total mass of the disk, 
$a$ and $b$ are scale parameters that 
control the radial and vertical extent of the disk  
respectively, and $R=\sqrt{x^2+y^2}$.
We investigate the models with $a=1.5$ kpc, $b=0.3$ kpc, and $M_{\rm d}=3 \times 10^{9}$ ${\rm M}_{\odot}$.
The combination of these parameters can give a reasonable and realistic
rotation curve for the LMC.
Since our major focus is to investigate how the LMC-SMC interaction about
2 Gyr ago influenced the formation processes of SCs in the SMC, we do not model the gravitational potential of the Galaxy; we have avoided introducing
unnecessary complication of the LMC-SMC-Galaxy dynamical modeling, despite this being required in the Magellanic Stream formation (\citealt{K3_Diaz_Bekki2012}). We mainly investigate the fiducial model, and the parameters are summarized in
Table \ref{table:1KParameters}.

\subsection{The SMC orbits}

\begin{figure}
 \includegraphics[width=\columnwidth]{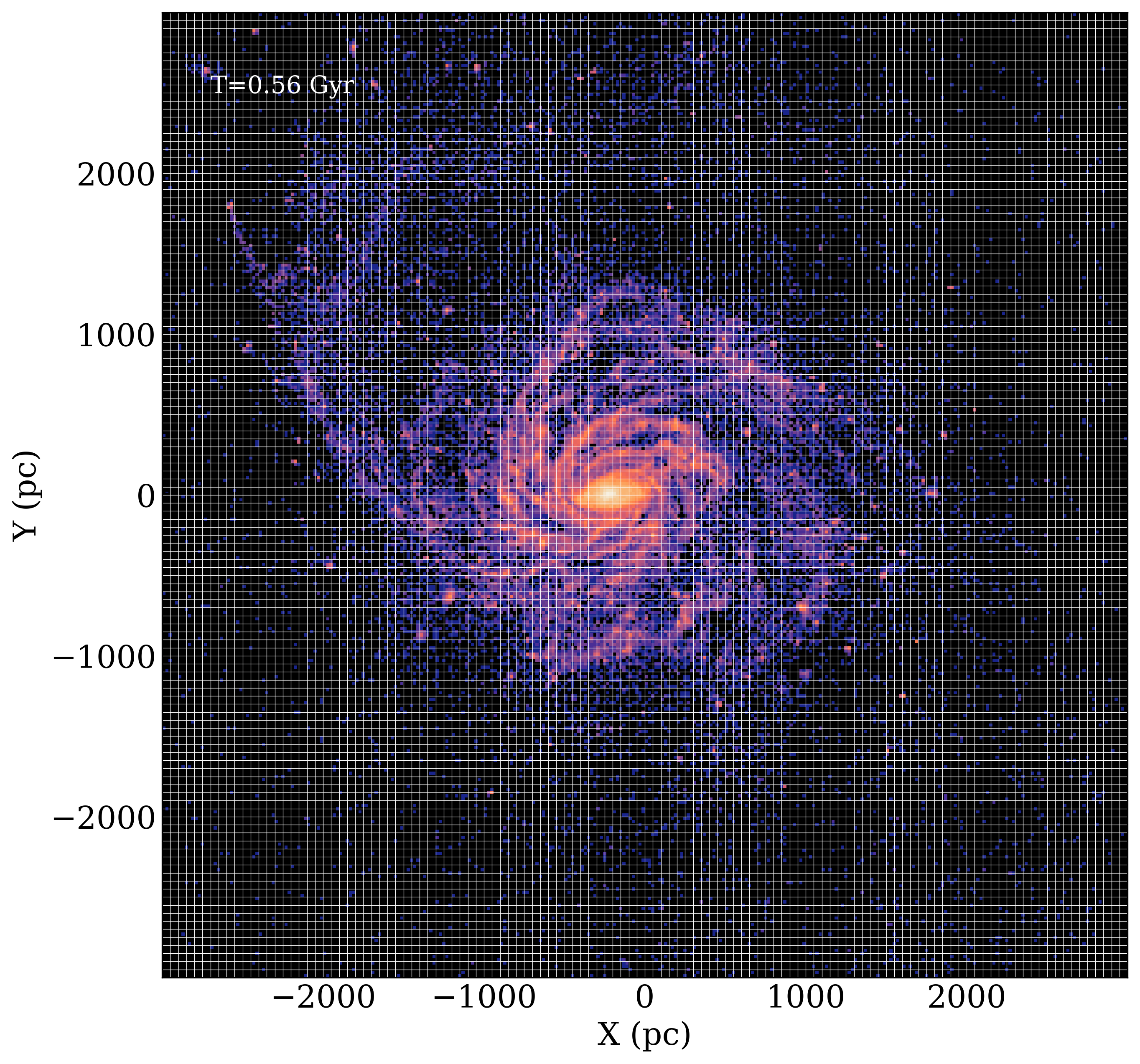}
 \caption{Two-dimensional (2D) map of surface mass density of new stars in the SMC projected onto the x-y plane at T=0.56 Gyr in the fiducial interacting model T1. The SMC disk is divided into $50 \times 50$ pc meshes, and thereby the mass density of new stars in each mesh is estimated to find the candidates of SCs.}
 \label{fig:3Mesh}
\end{figure}

We investigate the orbital evolution of the SMC around the LMC for ${\sim} 0.85$ Gyr in order to investigate whether the LMC-SMC interaction can induce the formation of massive SCs in the SMC disk, and how it does so. 
We consider that the SMC is initially located at it apocentre distance in all of our simulations, and therefore, the SMC has only the tangential velocity component ($V_y$); $V_x=V_z=0$. $Z_{\rm smc}$, which is fixed at 3 kpc, meaning that the SMC's orbit plane is close to the LMC disk.

Since we adopt the axisymmetric gravitational potential of the LMC, we consider that the $y$-component of the velocity of the SMC
is the most important free parameter for the SMC orbit. 
We investigate the models with $V_{\rm y}=30$, and 50 $\text{km}\text{s}^{-1}$ in the present study.
The larger $V_{\rm y}$ means a more circular orbit for the SMC. The SMC disk is inclined by 30 degrees with respect to the orbital plane of the SMC in all models of the present study.
\subsection{Simulating SC formation from GMCs}
\subsubsection{First step: SC formation sites}

In this first step, we investigate whether or not massive stellar clumps
can be formed during, and after the intense LMC-SMC interaction about 2 Gyr ago. In the adopted star formation model,
one gas particle is converted into one new stellar particle.
Therefore, one stellar clump
can consist of only ${\sim} 100$ new stellar particles, which is far
smaller than the typical number for massive SCs (${\sim} 10^5$): this first
analysis does not enable us to discuss the formation processes of SCs.
In order to identify such massive clumps we adopt a similar method as used in \cite{M7_McKenzie_Bekki_2021}. We divide the SMC disk with 3 kpc radius into $120 \times 120$ meshes and thereby investigate the total
mass of new stars at each mesh point (Fig. \ref{fig:3Mesh}). Next, we investigate the mass function of these clumps in order to discuss whether or not it can be influenced by the LMC-SMC tidal interaction. Furthermore, we 
investigate the number of massive clumps with masses more than 
$10^5$ ${\rm M}_{\odot}$
that can be regarded as massive SC candidates. 
We compare these mass functions between the interaction model (T1) and the isolated
one (I1) in which no LMC is included. We also investigate different SMC and LMC properties to see the effect on GMCs and new star formation. These parameter changes are outlined in Table \ref{table:1SMCmod}.

\begin{table}
\centering
 \caption{Parameter dependence of further interacting models. Both I1 and T1 have the same SMC properties.}
 \label{table:1SMCmod}
 \begin{tabular}{lccccc}
  \hline
  Model & LMC & SMC & Velocity & Pericentric & Tidal \\ 
                ID & mass & mass &V$_y$ & distance & radius\\
                &(M$_\odot$)&(M$_\odot$)&(km s$^{-1}$)&(kpc)&(kpc)\\
  \hline
  I1 & $6 \times 10^{10}$ & $6 \times 10^9$ & $30$ & $35.2$ & $11.7$\\ 
  T1 & $6 \times 10^{10}$ & $6 \times 10^9$ & $30$ & $20.2$ & $7.1$\\ 
  T2 & $6 \times 10^{10}$ & $1.2 \times 10^{10}$ & $30$ & $20.7$ & $9.2$\\
  T3 & $1.2 \times 10^{10}$ & $6 \times 10^9$ & $30$ & $27.2$ & $16.7$\\ 
  T4 & $1.2 \times 10^{10}$ & $6 \times 10^9$ & $50$ &$21.2$ & $12.6$\\ 
  T5 & $6 \times 10^{10}$ & $6 \times 10^9$ & $50$ &$23.1$ & $7.9$\\ \hline
 \end{tabular}
\end{table}

\subsubsection{Second step: SC formation processes}

In this second step, we investigate the mean densities of molecular hydrogen for all SPH particles at each time-step, and find a "high-density GMC" which has more than a threshold total mass ($M_{\rm th}$)
within 15pc for SC formation.
The threshold mass is a parameter to find a SC-forming GMC with a particular total mass range.
Once such a high-density GMC is found, then gas particles in the GMC
are converted into new stars. In this second step, one gas particle is regarded as one small gas cloud that can form many stars. Accordingly,
one gas particle is converted into $N \approx 1000$ stars so that the GMC consisting of $>100$ gas particles can form a SC with more than $10^5$ stars.
In this second step, once one high-density GMC is found at a time-step $T_{\rm f}$, the other high-density GMCs formed after $T_{\rm f}$ do not form stars in this new way, but with each gas particle being converted into one new star (as in the first step). This means to say that the simulation runs in its entirety each time, but with one high resolution cluster for analysis.
We adopt this model for the formation of just one "real" SC in this second step, because we need to avoid a huge number of stellar particles ($>10^7$) that could possibly form in many SCs, which is almost impossible
and much less practical to simulate in a reasonable
time scale. To avoid spurious numerical effects of having two different resolutions in the same simulation as much as possible, we allocated a much smaller gravitational softening length and time step width.

Each gas particle in the selected SC-forming GMC is assumed to be a small
spherical molecular cloud with a size of $h_{\rm g}$, where $h_{\rm g}$ is the
smoothing length of the SPH particle. Therefore,
each small cloud is converted into a small spherical cluster with
a size of $h_{\rm g}$ and a uniform spatial distribution within $h_{\rm g}$. The gravitational softening length is introduced for these new stars and is fixed at 0.3 pc. The maximum time-step width ($\delta t_{\rm max}$)  of these stars is smaller than those of other particles by a factor of 64 to 128, so that the dynamical evolution of the stars can be investigated.
For most models, $\delta t_{\rm max}$ is set to be $1.8 \times 10^4$ yr for these new stars in small clusters, which is much smaller than one dynamical
timescale for massive SCs.
Thus, this second step simulation allows us to investigate (i) whether 
a bound SC can be really formed in the SMC,
and (ii) what shapes, structures, and kinematics the SCs have.

\begin{figure}
 \includegraphics[width=\columnwidth]{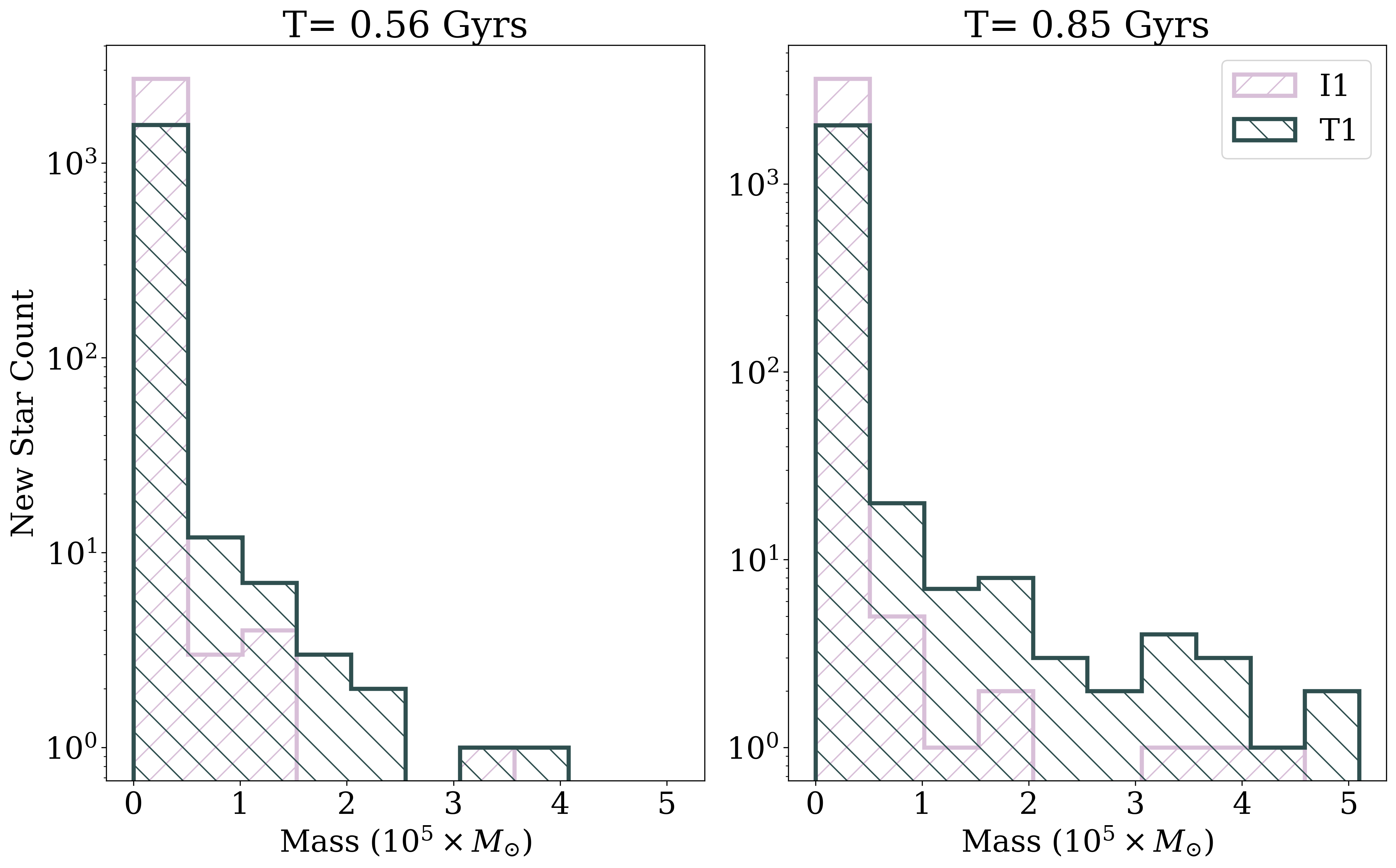}
 \caption{Histograms of the cumulative total masses of new stars throughout the simulation in the SMC at two different time-steps, T=0.56 Gyr (left) and 0.85 Gyr (right) for the tidal interaction model T1 (teal) and isolated model I1 (purple).
 The mass bin size is set to be $5.1\times10^4$ M$_\odot$ in these histograms.}
 \label{fig:4Histogram}
\end{figure}

\begin{figure}
 \includegraphics[width=\columnwidth]{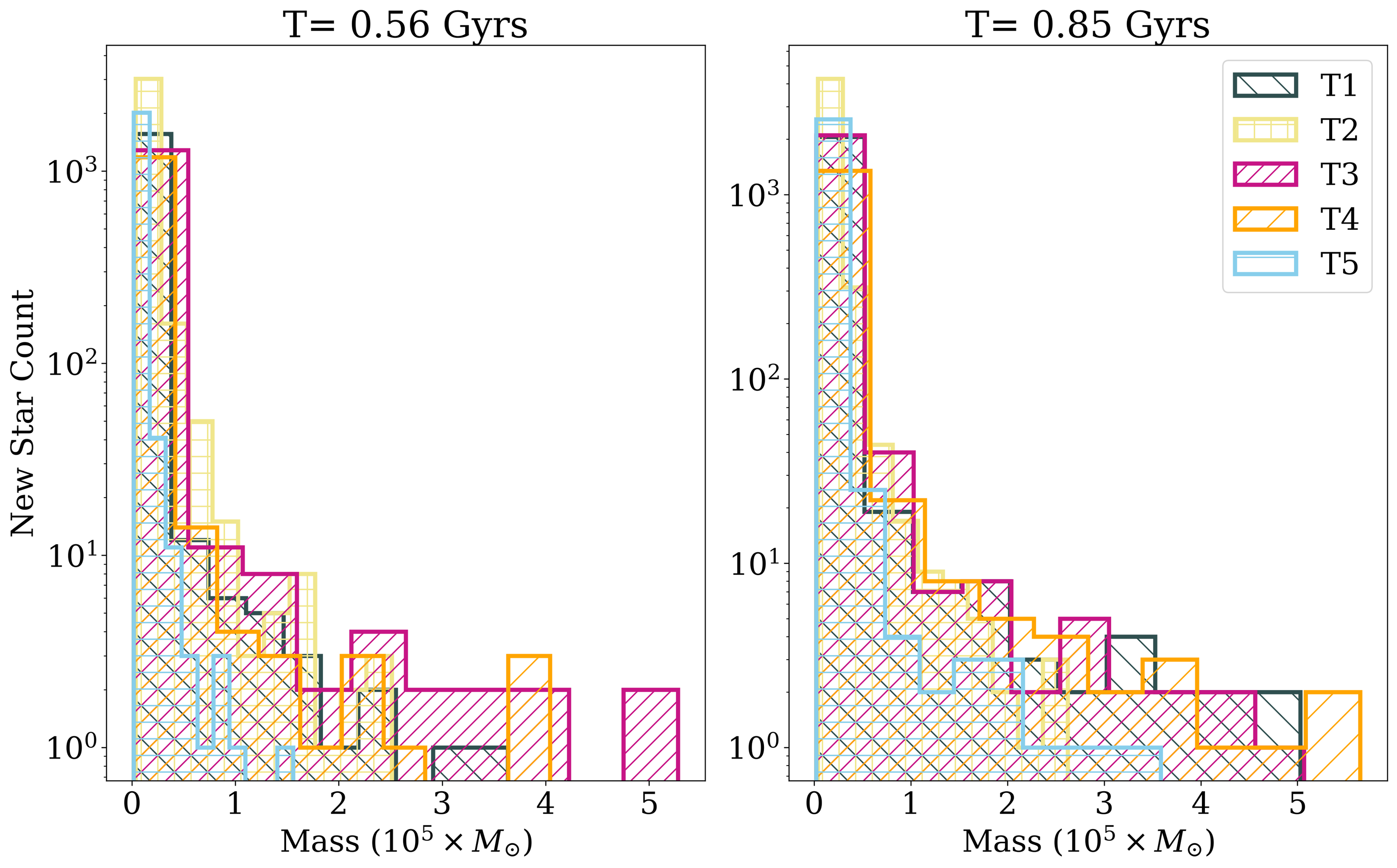}
 \caption{The same as Fig. \ref{fig:4Histogram} but for all tidal models given in Table \ref{table:1SMCmod}. Different colors indicate different models, and each model has 10 bins within the entire range of values to show data more clearly.}
 \label{fig:5Histall}
\end{figure}

\begin{figure*}
 \centering
 \includegraphics[width=15cm]{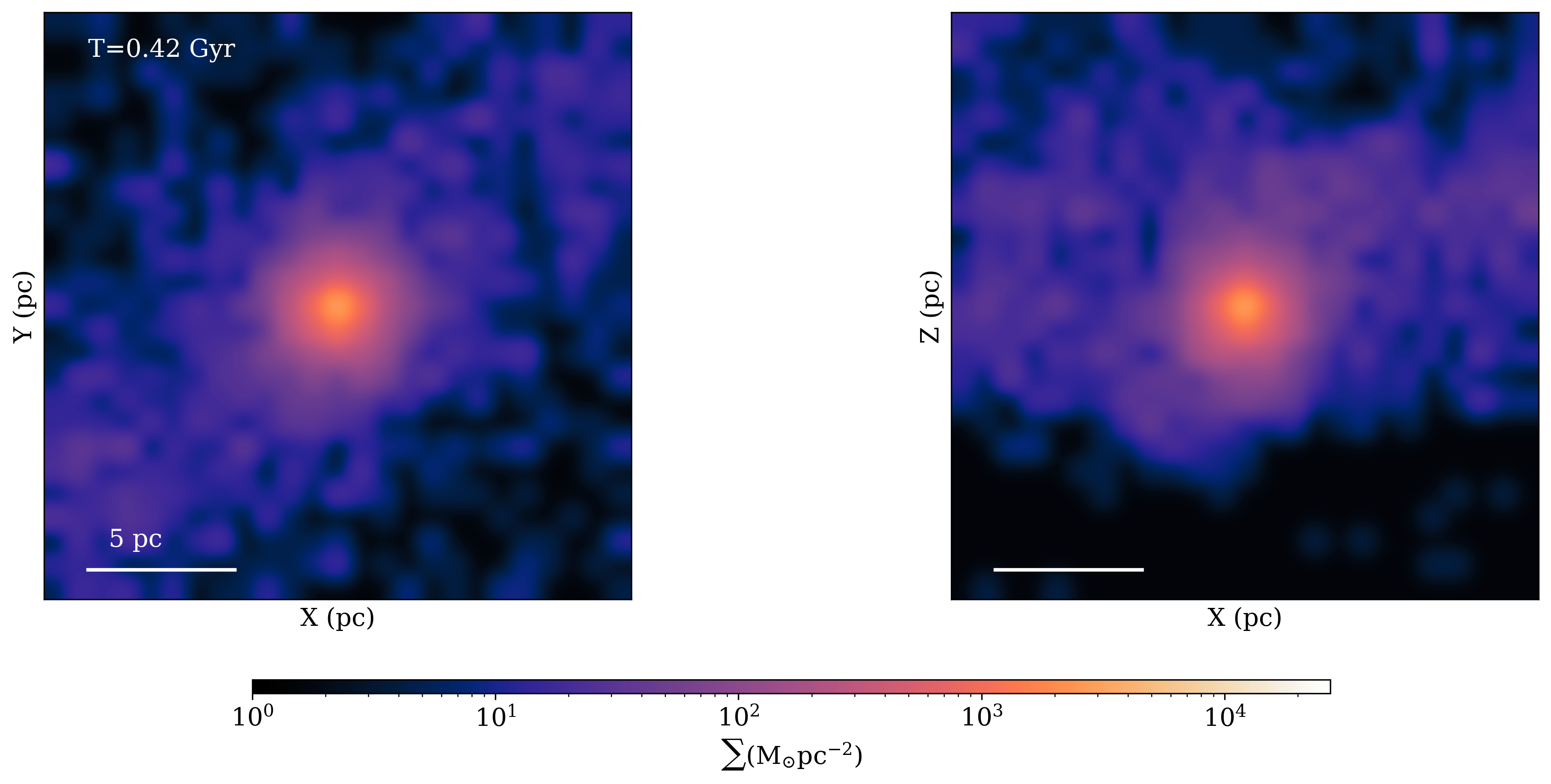}
 \caption{2D maps of surface density of new stars projected onto the x-y (left) and the y-z (right) planes at T=0.42 Gyr for SC1 in the tidal interaction model T1. This cluster shows more spherical shapes in the two projections, though other SCs show diverse morphologies of new stars.}
 \label{fig:6Cluster}
\end{figure*}

\begin{figure}
 \includegraphics[width=\columnwidth]{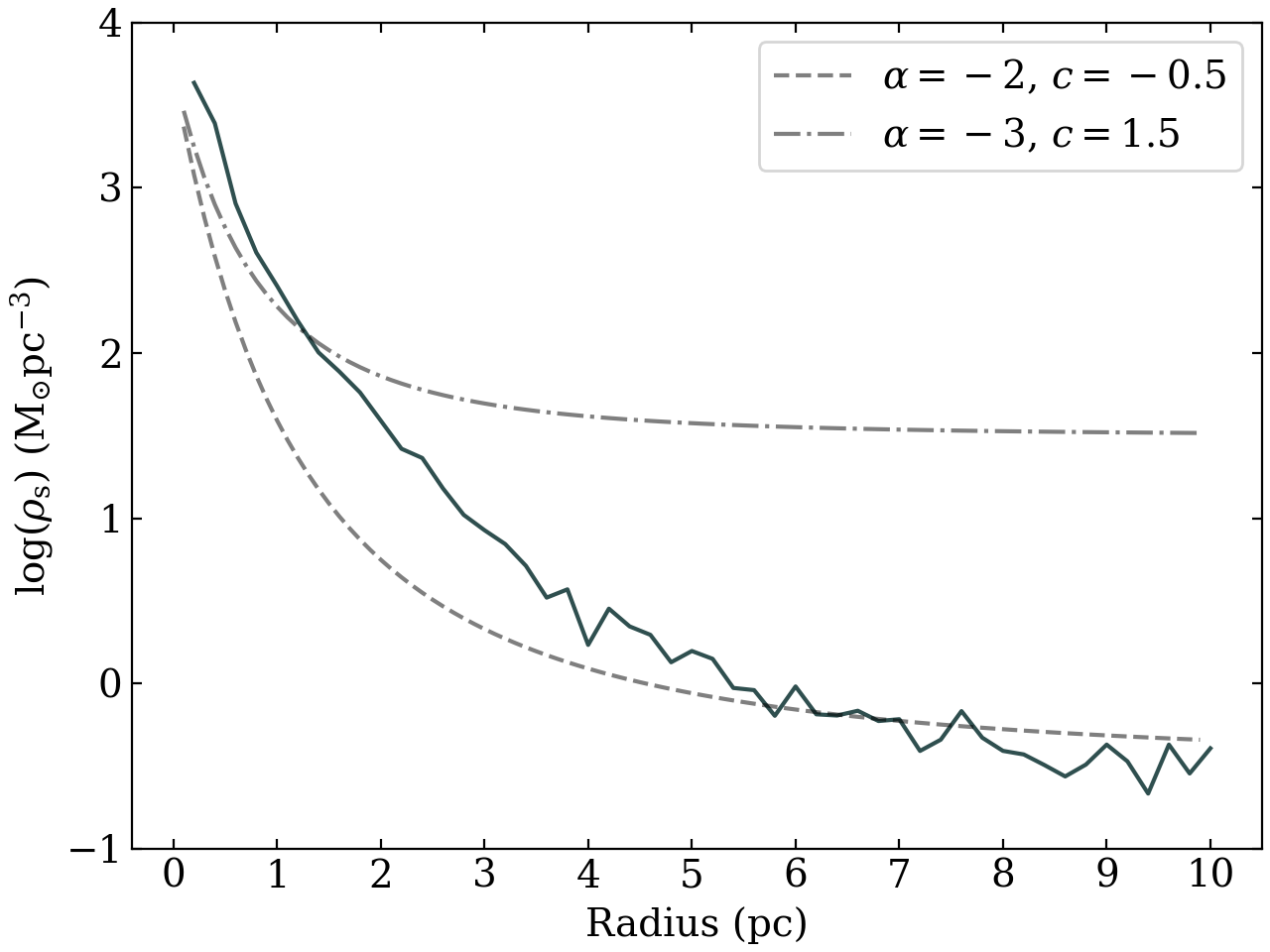}
 \caption{Radial density profile of SC1 with a total mass within $5$ pc of $8.4\times 10^3 \text{M}_\odot$ and radius ${\sim}3$ pc. The grey lines show 2 different power laws, with the dashed line having $\alpha =-2$ and $c=-0.5$, while the dashed-dotted line has $\alpha=-3$ and $c=1.5$.}
 \label{fig:7RadDense}
\end{figure}
We consider that the star formation efficiency within a small molecular cloud
depends on various physical parameters of the cloud, such as the size,
mass, density, and magnetic field strength etc. However, our simulations
cannot predict $\epsilon_{\rm sf}$, because one particle is allocated 
for the cloud. Therefore, we assume that $\epsilon_{\rm sf}$ is a
parameter ranging from 0.3 to 1.0. Since we choose only high-density GMCs
as SC-hosting gas clouds,  $\epsilon_{\rm sf}$ should be quite high.
Accordingly, we mainly investigate the models with $\epsilon_{\rm sf}=0.5$. We investigated the models with $0.3<\epsilon_{\rm sf}< 1.0$ and found that
the results do not strongly depend on $\epsilon_{\rm sf}$ as long as $\epsilon_{\rm sf}$ is larger than
0.5. However, the models with $\epsilon_{\rm sf}\geq 0.8$ (which is a bit too high) shows compact SCs with higher central densities. We do not discuss further about how the present results depend on this parameter $\epsilon_{\rm sf}$ in this study.

\section{Results}
\label{sec:results}
\subsection{Star cluster formation sites}
\subsubsection{Fiducial model}

\begin{table}
\centering
 \caption{Star clusters and their properties, with Mass (of new stars) and Half mass radius being taken within $5$ pc.}
 \label{table:2SCall}
 \begin{tabular}{lcc}
  \hline
  Cluster ID & Mass ($\text{M}_\odot$) & Half Mass Radius (pc)\\ \hline 
  SC1 (fiducial) & $8.39\times10^3$ & 1.27\\
  SC2 & $9.48\times10^4$ & 1.59\\ 
  SC3 & $3.77\times10^4$ & 2.05\\ 
  SC4 & $2.04\times10^3$ & 1.41\\ 
  SC5 & $1.25\times10^4$ & 3.35\\
  SC6 & $2.06\times10^4$ & 3.09\\
  SC7 & $1.20\times10^5$ & 2.48\\
  SC8 & $5.53\times10^4$ & 1.10\\
  SC9 & $1.82\times10^4$ & 2.54\\ \hline
 \end{tabular}
\end{table}

We investigate the influence of tidal interactions with an LMC-like potential on our SMC models, with Table \ref{table:1SMCmod} describing five different models we tested during our study. We present the results to our fiducial model, T1, in the remainder of this section.

Firstly, we investigate how the gas disk of the SMC evolves with time during its violent tidal
interaction with the LMC. Fig. \ref{fig:1SMCall} shows the time evolution of the surface mass density of the gas disk in the SMC for the fiducial tidal interaction model T1. 
In this model, the initial exponential disk is inclined on a 30$^\circ$ with respect to the x-y plane.
From the initial conditions outlined in the previous section, the dwarf galaxy undergoes an interaction with our LMC-like potential between the second and third time frame, causing it to become a rotating disk with a barred centre. Analysing the distance between the SMC and the centre of the LMC in the simulation, the interaction has been caused by a close encounter between the LMC and the SMC at ${\sim}0.3$ Gyr. This explains the change in the simulation after this time-step.

\begin{figure*}
 \centering
 \includegraphics[width=15cm]{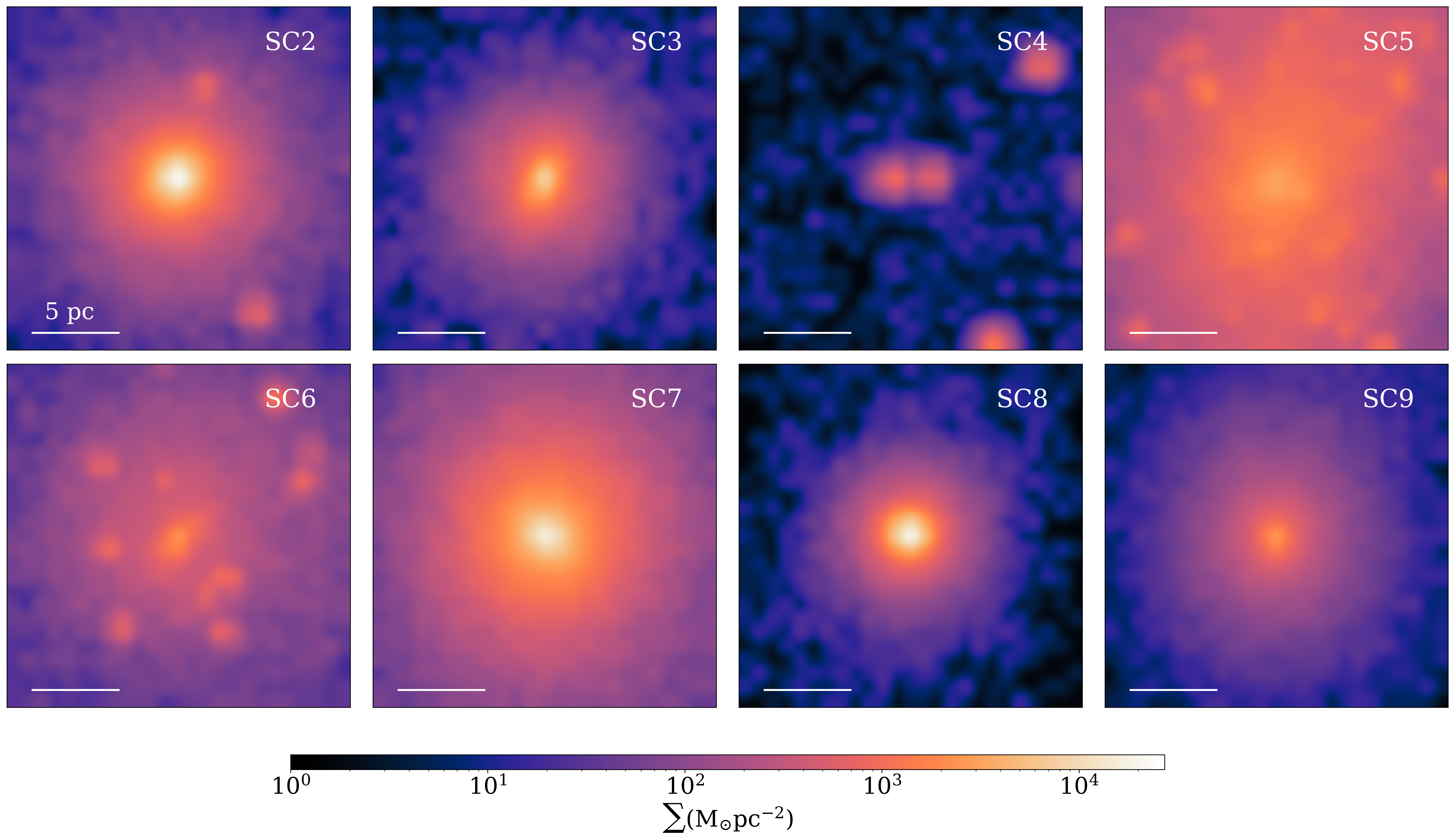}
 \caption{2D maps of surface density of new stars projected onto the x-y plane for 8 SCs (SC2 to 9) at T=0.42 Gyr. Clearly, the projected shapes of these SCs are quite diverse ranging from central massive clusters with surrounding sub-structures, to binary SCs, and to multiple sub-clusters.}
 \label{fig:8Clusterall}
\end{figure*}

\begin{figure}
 \includegraphics[width=\columnwidth]{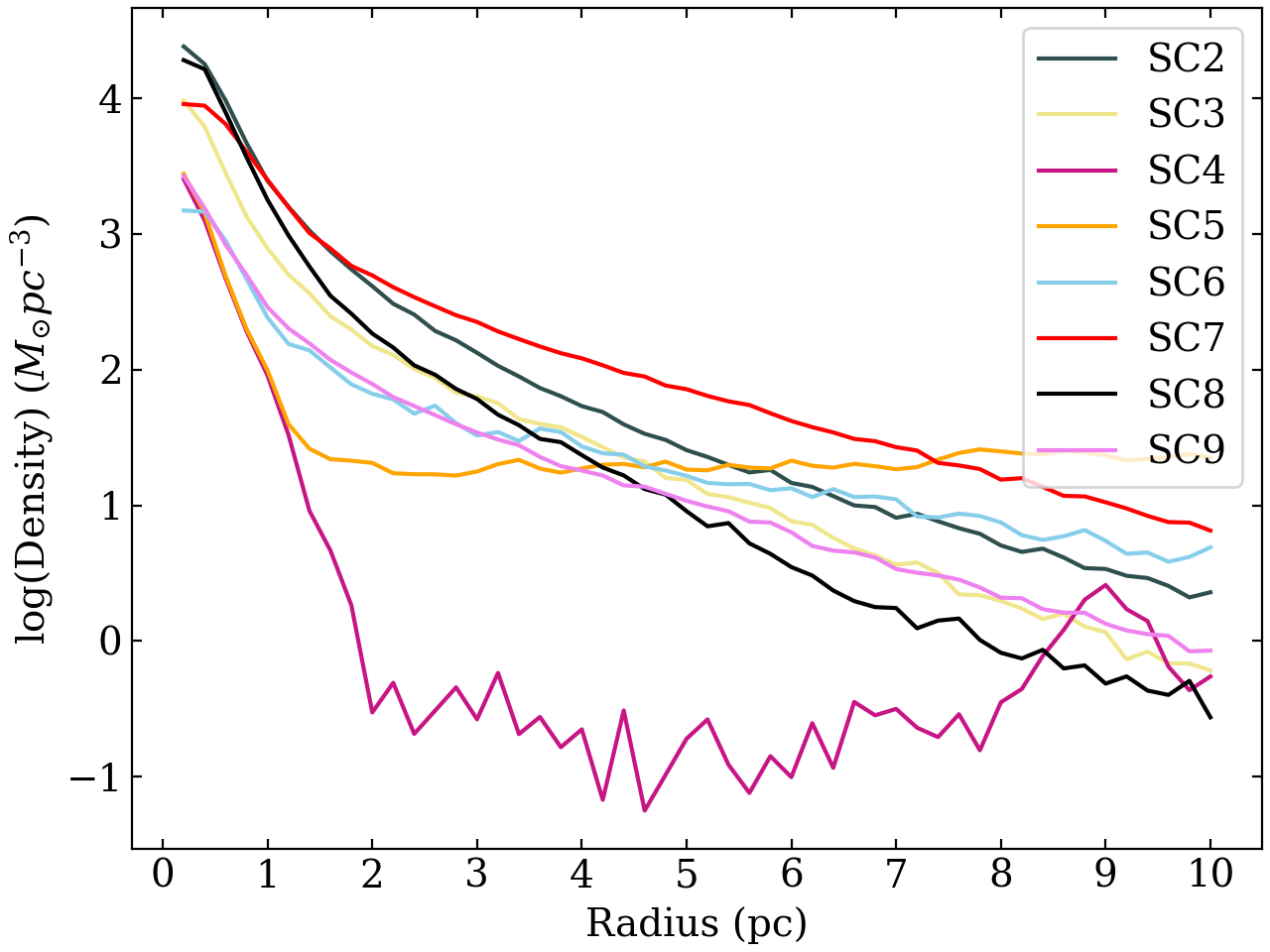} 
 \caption{Three-dimensional radial density profiles of 8 SCs (SC2-9). Different colors indicate different SCs.}
 \label{fig:9RadAll}
\end{figure}

There is also a notable high-density region forming in the centre of this model in the final two time frames. To investigate this, we look at the star formation rate (SFR) over the whole simulation, as SFR is proportional to local gas density. 
Fig.~\ref{fig:2SFR} shows the SFR over 0.8 Gyr of the simulation, where both models start with a SFR of ${\sim} 0.02 \rm{M}_{\odot} \ yr^{-1}$, however, after ${\sim}0.6$ Gyr in the simulation, the SFR of T1 increases by a factor of $2.3$ in comparison to I1.
In I1, there is no increase in SFR, and hence showing the increase in local gas density due to the tidal interactions. This enhancement factor aligns with \cite{21Bergvall_Laurikainen_Aalto2003}, where an enhancement factor of 2-3 was also noted.\newline

\begin{figure*}
 \includegraphics[width=15cm]{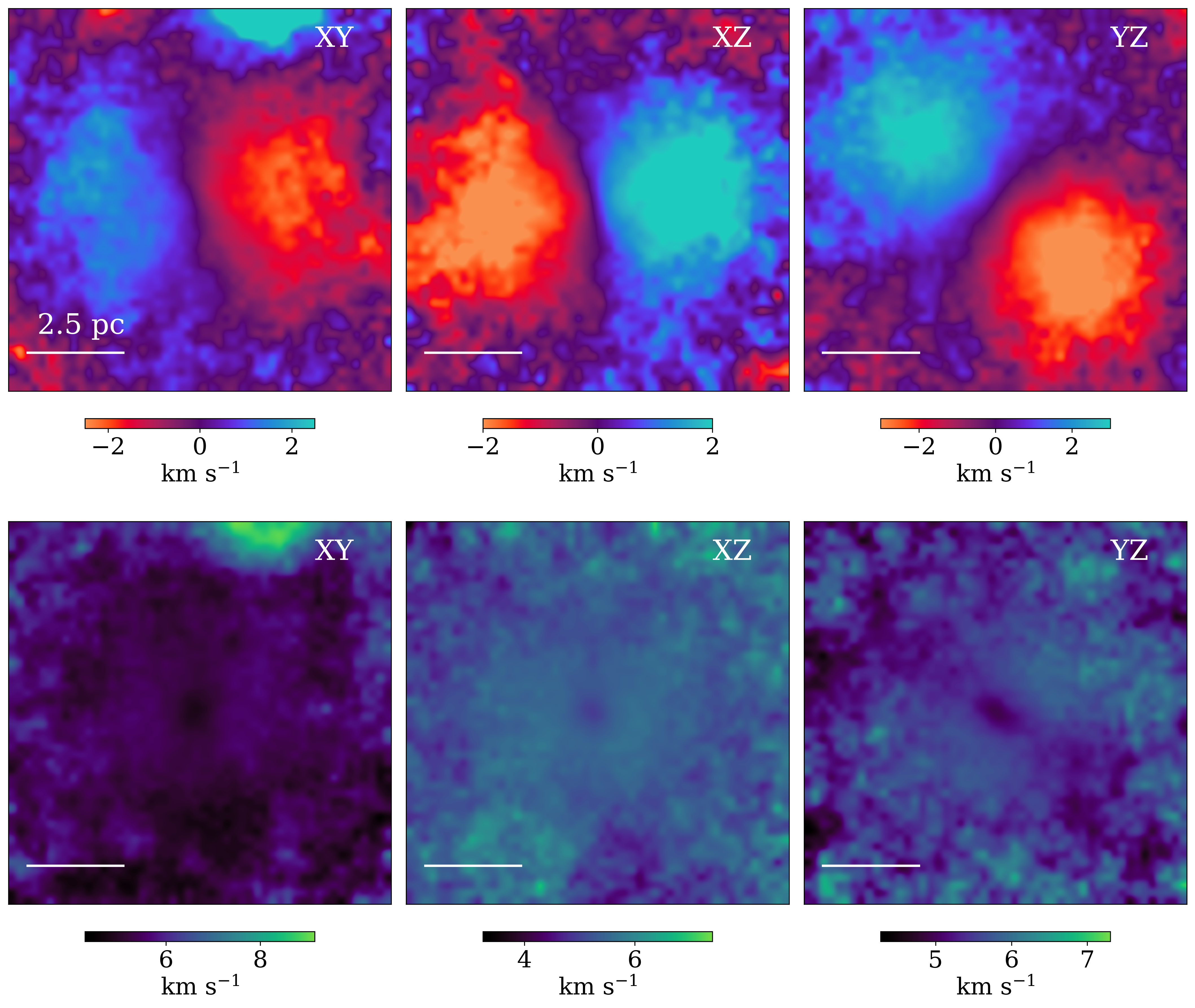}
 \caption{2D maps of line-of-sight velocities centred on the average velocity (upper row) and velocity dispersion (lower row) of new stars projected onto the x-y (left), x-z (middle), and y-z (right) planes for SC2.
 Dynamical relaxation acts on much longer timescales than what has been simulated, therefore these 2D kinematics maps show irregular patterns in the outer stellar envelope. However, it is clear that SC2 has global internal rotation in the three projections, which implies that this SC obtained internal rotation during its formation through hierarchical merging of sub clusters formed from their natal GMC.}
 \label{fig:10Rotation}
\end{figure*}

Further analysis was done to extrapolate potential cluster regions. We expected the clusters to have a radius between 5-15 pc  (\citealt{4Boutloukos_Lamers2003}; \citealt{11Bitsakis_et_al2018}), so a mesh with squares size $50\times 50$ pc was calculated for each time frame, and the total mass of new stars 
was calculated for each mesh. Fig. \ref{fig:3Mesh} gives an example of this process, which shows this 120$\times$120 mesh projected onto the T = 0.56 Gyr time-step. For this research, the mesh was applied in the x-y plane. We did investigate taking the mesh in the SMC's inclined plane, however this produced the same enhancement factor, and given that the clusters would be re-simulated spherically in the second stage, the choice between these orientations didn't effect our results. The distribution of masses in each bin is then presented in Fig. \ref{fig:4Histogram} for both T1 and I1. As a result of the increased SFR, T1 has more massive stellar clumps (with over $2\times 10^5$ $\text{M}_{\odot}$) than I1 in both the T = 0.56 Gyr and 0.85 Gyr time-steps, further demonstrating the enhancement of potential star cluster regions.

\begin{figure*}
 \includegraphics[width=15cm]{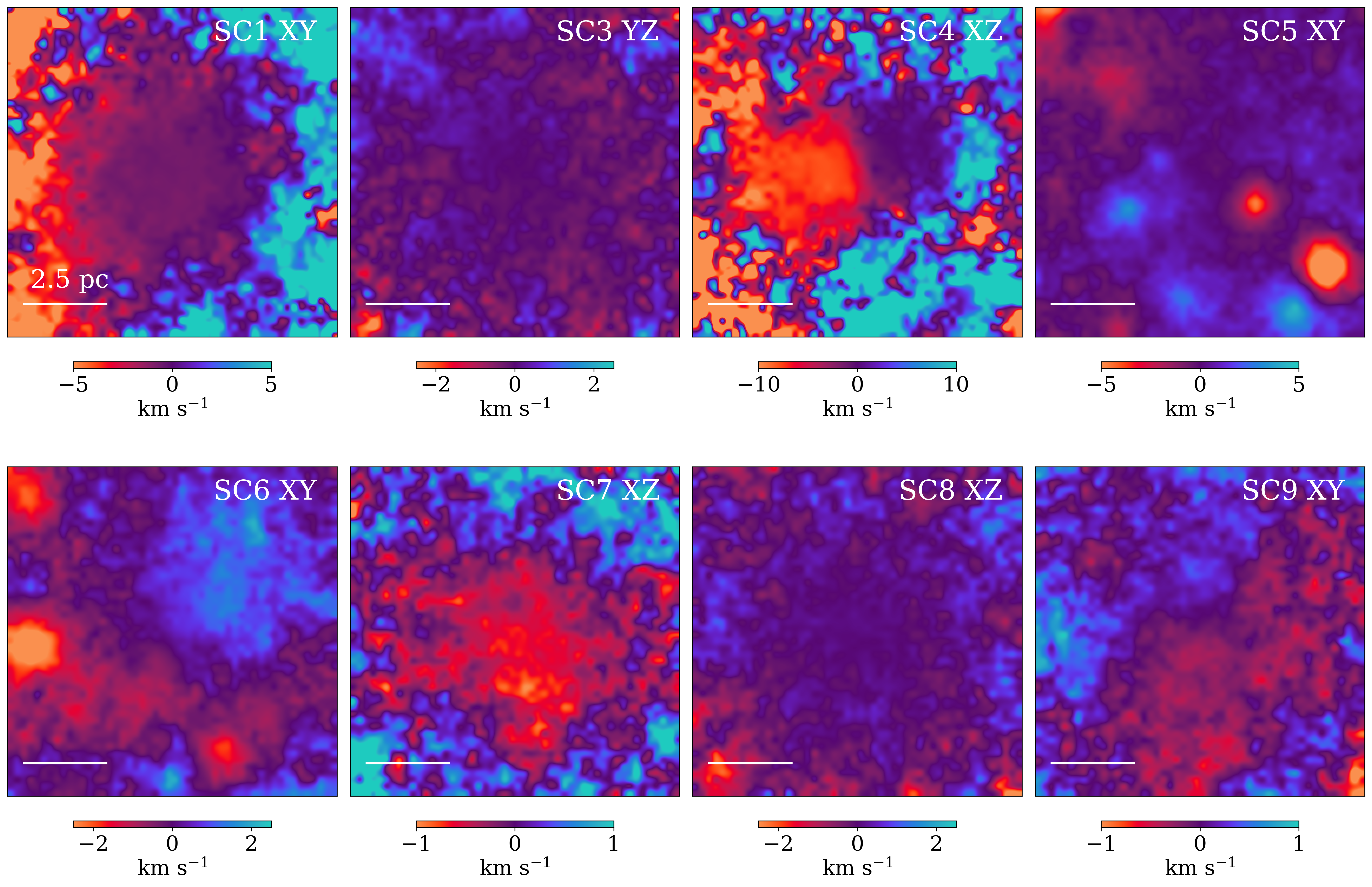}
 \caption{2D maps of line-of-sight velocities centred on the average velocity for 8 selected SCs. The viewing angles for these 8 SCs are chosen such that characteristic stellar kinematics can be seen more clearly.}
 \label{fig:10.5velocity}
\end{figure*}

\begin{figure*}
 \includegraphics[width=15cm]{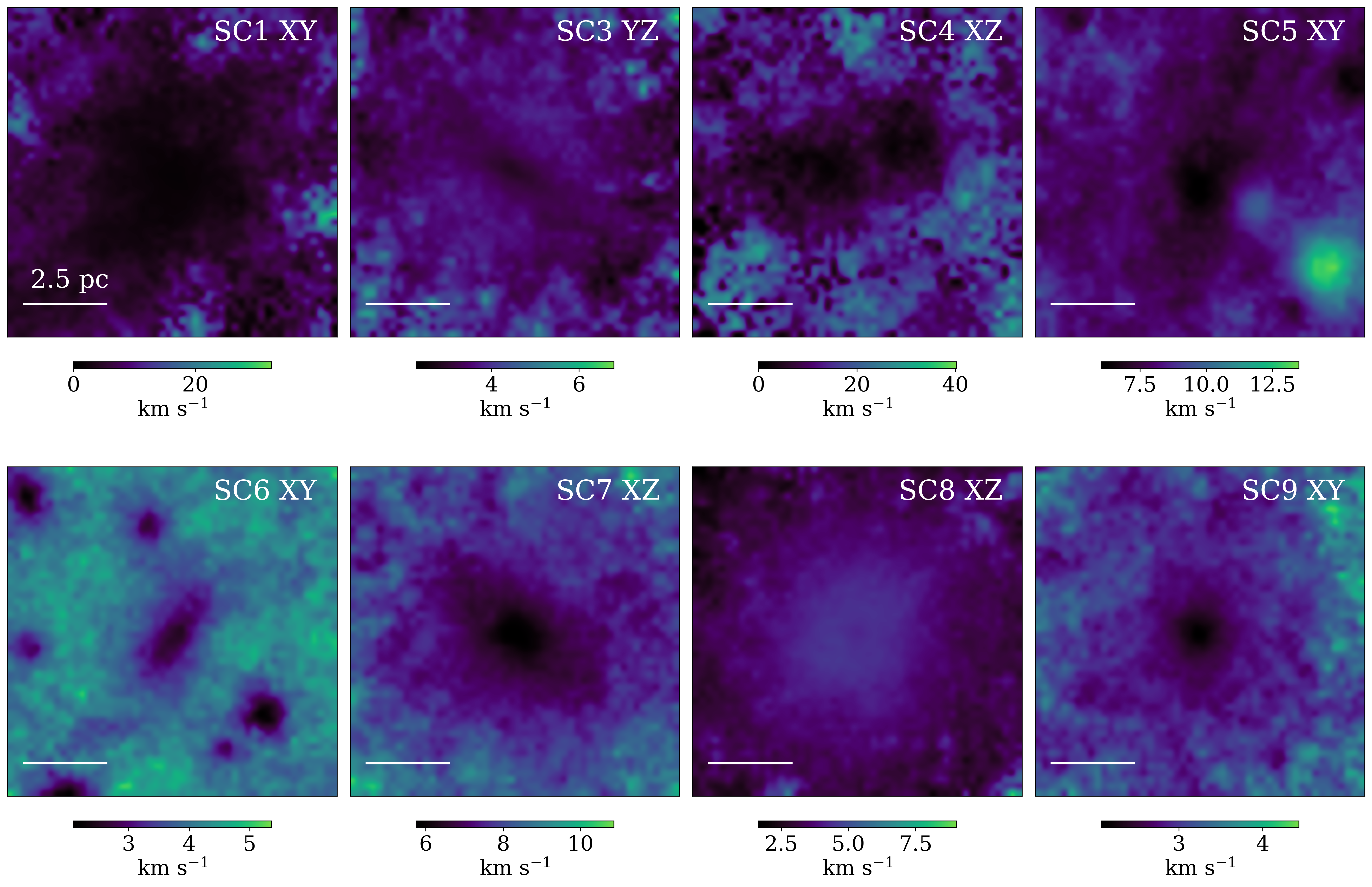}
 \caption{The same as Fig. \ref{fig:10.5velocity} but for velocity dispersions.}
 \label{fig:10.6dispersion}
\end{figure*}

\subsubsection{Parameter dependence}

There were four other models investigated, all with varying parameter dependence (Table \ref{table:1SMCmod}). T2 has a noticeably higher SFR, as expected from the larger SMC model mass. Despite T3, T4 and T5 following the trends shown by T1, T2 shows no predominate SFR increase after the tidal interaction (Fig. \ref{fig:2SFR}). This is also seen in the mass function (Fig.~\ref{fig:5Histall}), 
 where T2 has no regions with density greater than $3\times 10^5$ ${\rm M}_{\odot}$. T5 shows an increase in SFR, however this occurs a lot later than T1, and, when looking at density regions, T5 has less high-density regions than I1. This could be explained by the circular orbit creating less high-density regions. T3 and T4 have very similar shapes to T1, however both form more high-density regions at earlier times than T1, with T4 reaching the highest density at ${\rm T}=0.85$ Gyrs of over $5\times 10^5$ ${\rm M}_{\odot}$. \\
 \\
 Despite there being recent works suggesting the LMC mass is greater than $1\times 10^{11}$ M$_\odot$ (\citealt{R3Pernarrubia_et_al2016}; \citealt{R1Erkal_et_al2019}; \citealt{R2Erkal_Belokurov2020}), we only investigated LMC masses up to $1.2\times 10^{10}$ M$_\odot$. For higher masses, we would expect a smaller tidal radius, which would imply the LMC has a larger influence on the SMC.
\\
\subsection{SC formation}
Here we select 9 SCs from the tidal interaction model T1 which were formed from GMC's with initial mass over $5\times 10^{5}$ M$_\odot$, and thereby describe the physical properties of the SCs. These SCs are selected, mainly because each of them have quite interesting and unique characteristics in the internal structures and kinematics. We first describe the radial density profiles, the 2D density maps for SC1, followed by a further 8 SCs. Then we describe the 2D kinematics maps of line-of-sight velocities and velocity dispersion in detail.

\subsubsection{SC1 properties}

SC1 can be identified as a cluster due to the distinct spherical shape as seen in Fig. \ref{fig:6Cluster}. The irregular and clumpy structure in the cluster's outer stellar envelope is a result of the outer part of SC1 not being fully dynamically relaxed yet. The radial density profile (Fig. \ref{fig:7RadDense}) has a flat outer profile, which indicates a potential stellar envelope past 5 pc. It can also modelled by the following (3D) radial density profile:
\begin{equation}
    \log(\rho_{s})=15\times{\Big(\frac{2.4+r}{1.27}\Big)}^{\alpha}+c
\end{equation}
where $\alpha=-2$ and $c=-0.5$. \cite{K31Elson_Fall_Freeman1987} shows that the outer 2D projected density profiles have the power law with the slope of $\gamma$ ranging from -2.2 to -3.2, so our results align. SC1 has a mass of $8.4\times 10^3$ ${\rm M}_{\odot}$ within a $5$ pc radius, and $1.27$ pc half mass radius (mass taken within $r<5$ pc). As the SCs originally have GMC larger than $2\times 10^5$ M$_\odot$ within 15 pc, most new stars exist outside the half-mass radius. 
SMC GC's have a mass of ${\sim}2\times 10^5$ ${\rm M}_{\odot}$ within $3$ pc, meaning the mass and size of SC1 is too small to be considered a GC (\citealt{85Hill_Zaritsky2006}), however the mass is much greater than $300$ ${\rm M}_{\odot}$ within $2$ pc that open clusters (OCs) have.
Hence we identify the cluster as a massive star cluster. 

\subsubsection{SC1 kinematics}
Although the 2D maps of line-of-sight velocities and velocity dispersion's of this SC are investigated for the three x-y, x-z, and y-z projections, no/little rotation is found in this SC. Also, it was found that this cluster did not show a flattened 2D velocity dispersion map, that is indicative anisotropy in velocity dispersion's. These results of stellar kinematics
are consistent with the spherical shapes of this SC shown in Fig. \ref{fig:6Cluster}. This SC shows a large range of the projected velocities (e.g., -15.9 to 13.4 km s$^{-1}$ for the x-y projection), which would make it harder for this study to detect a small amplitude of rotation. It is, however, confirmed that this SC does not show any clear rotation in the three projections, even if the velocity range is limited to -5 km/s to 5 km/s. This point is discussed in Appendix A in detail, because careful analysis is required for the detection of internal rotation of SCs in the present study.

\subsubsection{SC2-9 properties}

It was found that SCs with diverse internal structures and kinematics can be formed from GMCs with different masses and kinematics in the tidal interaction model T1. Fig. \ref{fig:8Clusterall} summarizes the projected 2D mass distribution of new stars in the other eight SCs (SC2-9) formed in T1. SC3 shows a binary SC, and SC5 and SC6 show multiple SCs, however these observations are based of the projected distributions, as we did not investigate the mutual motions of these sub clusters. The clumpy nature of these clusters are due to the violent hierarchical merging of sub-clusters formed in their natal GMCs. Some sub-clusters cannot be totally destroyed during such merging events and can still stay in the outer parts of the central regions. These survived sub-cluster populations can form the clumpy distribution of stars around these SCs. Fig. \ref{fig:9RadAll} shows that the radial density profiles are significantly different between these 8 SCs, which can reflect the diverse formation histories and physical properties of their original GMCs. For example, the central density of the SCs range from 17.3 M$_\odot$pc$^{-3}$ to 650 M$_\odot$pc$^{-3}$. The flat outer profile is most common, as seen in SC2, SC3, SC6, SC7, SC8 and SC9. This shape is like that of SC1, implying a stellar envelope surrounding these clusters; a feature which has been observed around some galactic GCs (\citealt{M1_Kuzma_Da_costa_Mackey2018}), and some very young SCs in the LMC (\citealt{K31Elson_Fall_Freeman1987}). Such diffuse stellar halos have not been so far observed in the older SCs in the LMC and the SMC, which implies that the outer halo has been lost by some physical processes such as tidal stripping by the two galaxies. 
Observationally, cluster memberships of these outer halos regions could be confirmed with high precision astrometry of these clusters' outer regions using satellites such as Gaia (e.g. \citealt{GaiaCollaboration_2021}). 
SC4 and SC5 may have most of these surrounding stars unbound, however this cannot be inferred from the profiles.

\begin{figure*}
 \centering
 \includegraphics[width=15cm]{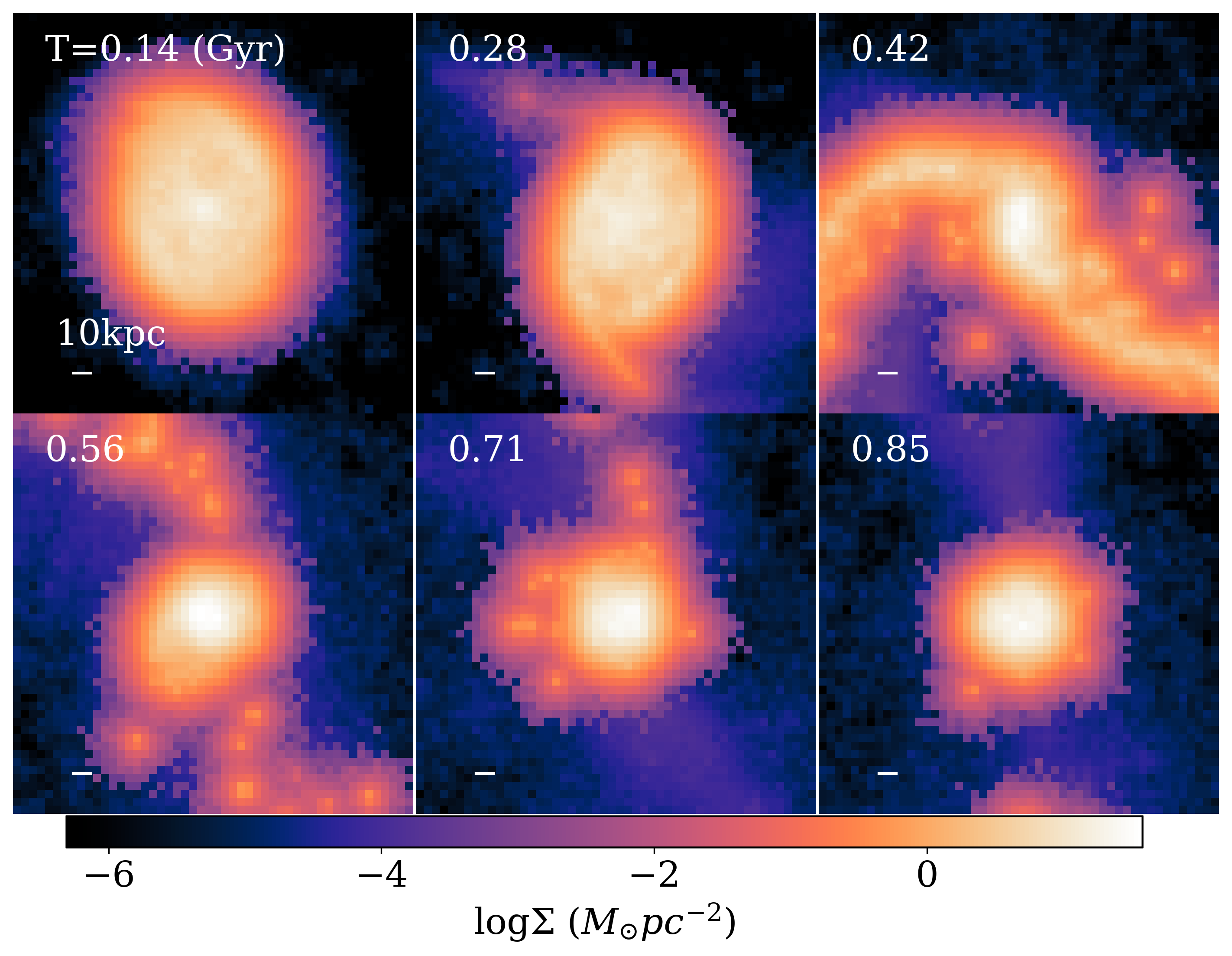}
 \caption{Time evolution of the 2D map of
H2 surface gas density  in the SMC during the LMC-SMC tidal interaction. Clearly, a ring-like distribution of molecular hydrogen can be
seen at T=0.28 Gyr.}
 \label{fig:11KenH2}
\end{figure*}

\subsubsection{SC2-9 Kinematics}
Two main properties of the clusters were investigated; their internal rotation and velocity dispersion. We first focus exclusively on the results of SC2, because it shows the most
intriguing behaviour in stellar kinematics. SC2 shows very clearly global rotation in the 2D line-of-sight velocity maps projected onto the x-z and y-z planes (Fig. \ref{fig:10Rotation}). The rotation amplitude can be as large as 9.2 kms$^{-1}$, with rotation visible over -5 < $x$ < 5 pc, and the cluster looks like a flattened spheroid or a bar. These results suggest that this cluster is dynamically supported largely by rotation. This high amplitude of global rotation in this cluster would be due to the larger angular momentum of its parent GMC. During the hierarchical merging of sub clusters in the early formation phase of this cluster, the orbital angular momentum of these sub clusters (i.e., rotating around the mass centre of the GMC) is converted into the intrinsic angular momentum of the cluster. The 2D velocity dispersion map shows the central dip, which indicates that there is a dynamically cold stellar disk in the SC. This inner disk is formed during gas accretion onto the SC from the GMC. The lower velocity dispersion confirms that the SC is supported by rotation rather than random motion of stars. It should be stressed here that global internal rotation of stars can be readily identified in this SC owing to the strong rotation. However, it is difficult for the present study to clearly identify such global rotation in some SCs from their 2D maps
of velocities, because the rotation amplitudes are significantly smaller compared to the velocity ranges of the 2D maps. This point is discussed in Appendix A, using the results of stellar kinematics derived for SC1.

In contrast SC3 does not show any clear rotation, which suggests that this SC has no or little global rotation within the central 5 pc. The total stellar mass of SC3 is by a factor of 3 times smaller than that of SC2, which implies a relation between SC masses and the amplitudes of global rotation. The velocity dispersion map for SC3 shows a cold flattened region (as shown in Fig. \ref{fig:10.6dispersion}) that is formed from later gas accretion within the SC and the subsequent star formation in a disk-like central region. It is beyond the scope of this paper how the flattened velocity dispersion map evolves with time due to long-term effects of two-body relaxation.

SC4 is a binary SC, and each of the two smaller SCs does not show any internal rotational stellar kinematics, only their rotation around each other (Fig. \ref{fig:10.5velocity}). SC5 has no noticeable features, besides its formation being later in the simulation, as evident by the range in line-of-sight velocity dispersion from 6.0 km s$^{-1}$ to 23.1 km s$^{-1}$. This shows the regions where gas particles are formed into many new stars. 
SC6 and SC9 show internal rotation with a small amplitude in the x-y plane, though the SC6 2D velocity map appears to quite irregular and clumpy. The 2D velocity dispersion map of SC6 also shows a flattened shape in its central region, which is due to efficient star formation in the central cold gas disk later formed in the forming GC.
SC7 and SC8 have no internal rotation, however SC7 and SC9 do have a central region with very low dispersion velocity, showing a cold central region.
These structures are in agreement with observations, noted in \cite{17Bekki_Chiba2009}, suggesting SMC clusters could be barred and have central rotation. 
\section{Discussion}
\label{sec:discussion}
\subsection{Is LMC-SMC interaction the best mechanism
for the enhanced SC formation ?}

As observed by \cite{1Rafelski_Zaritsky2005}, there is more than one peak of SC formation. The peak from $0.2$ Gyr onward (e.g., \citealt{27Grebel_et_al1999}, \citealt{40Glatt_Grebel_Koch2010}, \citealt{11Bitsakis_et_al2018}) is furthermore consistent with an interaction between the MCs (e.g., \citealt{17Bekki_Chiba2009}). However whether this was a tidal interaction or a collision is not agreed upon. \cite{11Bitsakis_et_al2018} 
investigated the age distributions of 1319 SCs in the SMC and suggested that a direct collision $0.1-0.3$ Myr ago would create the desired peak in SCs. They also proposed that such a recent LMC-SMC collision can explain the observed kinematics of the LMC stars that were suggested to be from the SMC (\citealt{18Olsen_et_al2011}). This would also explain the higher enhancement factor, as \cite{23Knapen_Querejeta2015} mentions mergers increase SFR the most. However, tidal interactions would enhance SC formation even without a collision (\citealt{79Bekki_et_al2004}), hence making our research relevant and potentially applicable regardless. 
In addition, there are also more papers suggesting this interaction would solely be a close encounter creating tidal interactions (e.g., \citealt{27Grebel_et_al1999}; \citealt{40Glatt_Grebel_Koch2010}; \citealt{39Nayak_et_al2018}), suggesting that this epoch of enhanced SC formation could also be from a tidal interaction, though this must be further investigated.

Another peak worth investigating from \cite{1Rafelski_Zaritsky2005} is the peak at ${\sim}2.5$ Gyr ago for Z$=0.008$ in Fig. 5 of their paper. This is approximately the same size as the $2$ Gyr peak, suggesting this could also be due to tidal interactions. Observationally, it might be very difficult to clearly distinguish between 2 Gyr and 2.5 Gyr stellar populations of SCs from color-magnitude diagram alone, despite them suggesting both a 2 Gyr and 2.5 Gyr peak. It would be possible that these two are indeed from a same burst population. Furthermore, two LMC-SMC interactions within 0.5 Gyr requires a strongly bound LMC-SMC orbit, which is not predicted by previous simulations of the LMC-SMC-MW interaction (\citealt{K15_Gardiner_Noguchi1996}, \citealt{K13_Besla_et_al2012}). We therefore consider that the two peak populations are likely to originate from a same starburst about 2 Gyr ago.

\subsection{Internal rotation of young SCs in the SMC}
We have shown, for the first time, that the young SMC SCs can have internal rotation just after their formation from their natal GMCs. 
Although this is one key prediction from the present simulation, almost all recent studies of SCs have investigated the internal kinematics of stars in the old Galactic GCs (e.g., \citealt{K21_Barth_et_al2020}; \citealt{K24_Kamann_et_al2020}; \citealt{K25_Szigeti_et_al2021}).
Accordingly, it is not possible for the present study to discuss the physical origin of the simulated rotational kinematics of some SCs based on the latest observations of SCs in the SMC. \cite{K18_Mackey_et_al2013} investigated the radial velocities of 21 stars in the LMC SC NGC 1846 and found a clear systematic rotation comparable to the velocity dispersion in this SC. Given the observed young age of this SC, it is likely that this SC initially had such clear rotation at its birth. Although this is not a result for the SMC SCs, this observation implies that other young SCs in the LMC and the SMC might have such global rotation within their stars.

Previous numerical simulations of SC formation on galaxy-scale and GMC-scale showed that SCs can have rotational kinematics just after their formation (e.g., \citealt{K26_Bekki2010}; \citealt{K27_Bekki2019}; \citealt{K28_Mapelli2017}). Although the long-term evolution of SCs with such rotational kinematics driven by two-body dynamical relaxation have been investigated by recent N-body simulations (e.g., \citealt{K29_Mastrobuono_Battisti_Perets2016}; \citealt{K30_Vesperini_et_al2021}), it is not fully understood how the rotational amplitudes of SCs can decrease due to dynamical evolution of compact stellar systems. The tidal field of the SMC and the tidal interaction of SCs with massive GMCs in the SMC can possibly influence the dynamical evolution of rotating stellar systems: it is possible that the 2 Gyr old SMC SCs do not clearly show large amplitudes of rotation even if they had at their birth, because the later long-term dynamical effects can dramatically decrease the original rotation of SCs. It is thus our future study to investigate the time evolution of stellar kinematics of SCs with initial rotation using an appropriate simulation code that implements both two-body stellar interactions and the (galaxy-scale) tidal effects by the SMC and its GMCs.

\subsection{Origin of sizes and masses of the SMC SCs}

\cite{11Bitsakis_et_al2018} calculates the mean SC half-mass radius in the SMC as $15.7$ pc using the measured distance modulus, with a minimum of $6.9$ pc and a maximum of $29.6$ pc. In comparison, the SCs discussed by \cite{4Boutloukos_Lamers2003} from visual inspection of photographic plates have a mean half-mass radius of $7$ pc, with a range from $5$ to $15$ pc. The SCs we simulated have a half-mass radii range of ${\sim}1-8$ pc from visual inspection. These values fall below both the range from \cite{4Boutloukos_Lamers2003} and \cite{11Bitsakis_et_al2018}. However further improvements must be made to our model to comment on the star formation fully, such as including more realistic feedback effects from massive stars with different masses (including supernovae), the statistical properties of SCs and the distribution of SCs within the SMC.

Observed SCs in the SMC have shown a correlation between their age and mass (e.g., \citealt{52Piatti2021}). \cite{52Piatti2021} also notes that smaller and younger clusters are found in the centre of the SMC. This aligns with our clusters being smaller than observed GCs with masses of ${\sim}3\times 10^5$ ${\rm M}_{\odot}$ (\citealt{85Hill_Zaritsky2006}). 

The SMC hosts one old, Galactic-like GC; NGC121. This cluster shows evidence of multiple stellar populations (\citealt{M5_Dalessandro_et_al2016}) and has a mass of $\approx 3 \times 10^5 \rm{M}_{\odot}$ (\citealt{M2_Mackey_Gilmore2003}).
However, due to its old age of $\approx$ 11 Gyr (\citealt{M3_Glatt_et_al2008}), we expect that it must have experienced some mass loss during its evolution and therefore, its initial mass must have been much higher. As our initial SC masses are smaller than the current mass of NGC121 (i.e. see Table 3) and we do not evolve our system long enough to test for cluster stripping and mass loss, we do not believe our SCs are large enough to be GCs. 

Alternatively, our clusters cannot be classified as OCs, as OCs have a mass of ${\sim}400$ ${\rm M}_{\odot}$, while our clusters are much larger than this. Hence, we can classify our clusters as massive star clusters. These clusters have been observed in \cite{52Piatti2021}, as having ages around $10^8-10^9$ yr old, and masses of ${\sim}10^{3.5}-10^4$ ${\rm M}_{\odot}$.\newline

\subsection{Formation of a molecular gas ring in the SMC}

Since our chemodynamical models include the formation of molecular hydrogen on dust grains, we can investigate the time evolution of the spatial distribution of molecular hydrogen in the SMC in the present study. Fig.~\ref{fig:11KenH2} shows that the intense LMC-SMC tidal interaction can form a ring-like distribution of molecular hydrogen between T=0.14 and 0.28 Gyr. Furthermore, a number of high-density small clumps, which corresponds to GMCs, can be seen in the ring: it is highly likely that star formation can proceed efficiently in the clumps. Such high-density clumps can be seen in the two tidal tails formed during the LMC-SMC interaction at T=0.42 and 0.56 Gyr too. Such an inhomogeneous
and clumpy distribution of molecular hydrogen disappears after the tidal interaction at T=0.85 Gyr, however, the SMC can finally have compact
molecular gas distribution with higher average gas density. This compact distribution is responsible for the constantly higher SFR in the SMC after the LMC-SMC tidal interaction.
\\
\\
The simulated ring-like distribution can be discussed in the context of the observed ring-like distribution of stellar populations with ages ranging from 1.6 Gyr to 2.5 Gyr (\citealt[HZ04]{20Harris_Zaritsky2004})
as follows. The simulated high-density molecular ring should be the place where star formation is very efficient, because SFR
is proportional to molecular gas density. Therefore, it is highly likely that the molecular ring ends up with the ring consisting of new stars only. If the ring-like distribution of new stars cannot be destroyed by the later LMC-SMC collision (${\sim}0.2$ Gyr) that formed the Magellanic Bridge (\citealt{K3_Diaz_Bekki2012}), then the ring-like structure should be able to be observed in the present SMC.
In the present study, we have only modelled the LMC-SMC interaction about 2 Gyr ago (not 0.2 Gyr ago), hence it is unclear whether or not the possible ring with 2 Gyr old stellar populations could survive during the violent dynamical heating of the SMC disk by the last LMC-SMC collision (0.2 Gyr). It is hence planned that future studies will investigate this issue using hydrodynamical simulations for the long term (more than 3 Gyr) dynamical evolution of the LMC-SMC system.

\subsection{Future works: from SC formation to
long-term evolution}

We have found, for the first time, that some
of the simulated SCs have diffuse stellar envelopes surrounding the main bodies, flattened inner structures, global rotation, and sub-structures in the halos at their birth. These structures can evolve significantly with time due to two body dynamical relaxation effects. However, we have not investigated the long term evolution of SCs (>1 Gyr) after their formation in the SMC. This  means that we cannot discuss a number of key issues related to the origin of SCs in the present study.

For example, it is not possible for the present study to investigate how the binary status of the simulated SCs can evolve over the next 2 Gyr (to the present time); will they merge to form a single SC, or not. We also cannot comment on how the simulated rotation can be less remarkable due to the two-body dynamical relaxation effects, and whether or not ISM can be accreted onto the new SCs to form new SCs, as discussed in recent papers (e.g., \citealt{K23_McKenzie_Bekki2021}). We need to improve our modeling in star formation within GMCs, including more realistic feedback effects from massive stars with different masses (including supernovae). Furthermore, we need to address other issues on SC formation in the SMC, which we do not discuss at all in this paper, such as the statistical properties of SCs (e.g., mass function) and the distribution of SCs within the SMC.

\section{Conclusions}
\label{sec:conclusion}

The observed burst of SC formation in the SMC $\sim$2 Gyr ago is one of the major problems in SC evolution in the SMC and LMC. In order to understand the physical origin of this, we have investigated the effects of tidal interactions between the SMC and LMC on SC formation ${\sim}2$ Gyr ago using our new numerical simulations with formation of GMCs and a two-stage SC formation model. We have also investigated, for the first time, the details of the internal structures and kinematics of the simulated SCs, such as 2D maps of velocities and velocity dispersions, in the SMC. Our findings were as follows:
\renewcommand{\labelenumi}{(\arabic{enumi})}
 \begin{enumerate}
    \item Tidal compression of cold molecular hydrogen induced by LMC-SMC interaction about 2 Gyr ago is the main physical reason for the formation of massive gas clouds (GMCs). 
    The tidal LMC-SMC interaction significantly impacts the mass function of GMCs when compared to the isolated model. There is a three times enhancement of meshes with mass over $4\times10^{5}$ M$_\odot$ in the final time step of the tidal model compared to the isolated model.
    
    \item Such GMCs can be formed in a number of LMC-SMC tidal interaction models, though the mass function of GMCs depends on the model parameters for the SMC orbits with respect to the LMC. The simulated massive and compact GMCs can be the major formation sites of SCs due to their high gas densities. These SC-hosting GMCs are mostly formed in the central region of the SMC in the present tidal interaction models.
    
    \item SCs have lower mass densities, with the average of all investigated SCs being 78 M$_\odot$pc$^{-3}$, which is comparable to those of the SMC or LMC SCs, but too low compared to those of the Galactic old GCs. 
    
    \item The structures of SCs are quite diverse, ranging from single SCs, to binary SCs, to multiple systems. The dynamically young SCs have stellar envelopes and sub structures around them in all models, which are not very similar to the fully developed older SCs in the SMC. The radial density profiles show the outer flatter component (corresponding to the stellar envelopes), which are similar to the observational results by \cite{K31Elson_Fall_Freeman1987}, which shows the projected density profiles with flatter outer parts compared to the truncated King model. These results imply that long-term dynamical evolution can shape up the dynamically young SCs.
    
    \item The 2D maps of line-of-sight velocities and velocity dispersion in the simulated SCs are also diverse, which provide important implications about the SC formation (for future observations of SC kinematics). One of the simulated SCs has a colder inner disk and barred centre with clear rotation, with the rotation maximum and minimum velocities as 9.2 km s$^{-1}$ and -5.4 km s$^{-1}$ respectively. Other SCs do not show clear internal rotation (having rotational amplitudes of less than 2 km s$^{-1}$). Binary SCs can be formed from a single GMC in some models. Dynamically cold central regions (indicated by lower central velocity dispersion) can be formed from later gas accretion of gas in forming SCs. Flattened shapes in 2D velocity dispersions, which are due to anisotropy of velocity dispersions, can be found in some models.

    \item These characteristic internal structures and kinematics such as stellar halos, global rotation, and flattened velocity dispersion maps in the simulated young massive star clusters can evolve with time owing to the dynamical evolution of SCs through two-body stellar encounters within a few Gyr. However, this important long-term dynamical evolution of SCs is yet to be investigated in the present study, because the adopted simulation code does not allow us to investigate the  two-body relaxation effects. It is thus our future study to investigate how such long-term dynamical evolution can change the internal structures and kinematics of young SCs in the SMC over a few Gyr.
 \end{enumerate}

\section*{Acknowledgements}

We would like to thank the anonymous referee for their kind feedback. This study would not have been possible without the ICRAR Summer Studentship program in 2020-2021. This study made use of the \textsc{PYTHON} packages \textsc{NUMPY} \citep{van2011numpy}, \textsc{SCIPY} \citep{2020SciPy-NMeth} and \textsc{MATPLOTLIB} \citep{matplotlibHunter:2007}. Our numerical simulations were run on the OzSTAR GPU supercomputer kindly made available to us at the Centre for Astrophysics and Supercomputing at the Swinburne University of Technology.

\section*{Data Availability}

The data underlying this article will be shared on reasonable request to the corresponding author.



\bibliographystyle{mnras}
\bibliography{Actual_Paper_Writing} 

\begin{thebibliography}{}
\makeatletter
\relax
\def\mn@urlcharsother{\let\do\@makeother \do\$\do\&\do\#\do\^\do\_\do\%\do\~}
\def\mn@doi{\begingroup\mn@urlcharsother \@ifnextchar [ {\mn@doi@}
  {\mn@doi@[]}}
\def\mn@doi@[#1]#2{\def\@tempa{#1}\ifx\@tempa\@empty \href
  {http://dx.doi.org/#2} {doi:#2}\else \href {http://dx.doi.org/#2} {#1}\fi
  \endgroup}
\def\mn@eprint#1#2{\mn@eprint@#1:#2::\@nil}
\def\mn@eprint@arXiv#1{\href {http://arxiv.org/abs/#1} {{\tt arXiv:#1}}}
\def\mn@eprint@dblp#1{\href {http://dblp.uni-trier.de/rec/bibtex/#1.xml}
  {dblp:#1}}
\def\mn@eprint@#1:#2:#3:#4\@nil{\def\@tempa {#1}\def\@tempb {#2}\def\@tempc
  {#3}\ifx \@tempc \@empty \let \@tempc \@tempb \let \@tempb \@tempa \fi \ifx
  \@tempb \@empty \def\@tempb {arXiv}\fi \@ifundefined
  {mn@eprint@\@tempb}{\@tempb:\@tempc}{\expandafter \expandafter \csname
  mn@eprint@\@tempb\endcsname \expandafter{\@tempc}}}

\bibitem[\protect\citeauthoryear{{Armstrong}, {For}  \& {Bekki}}{{Armstrong}
  et~al.}{2018}]{K33_Armstrong_et_al2018}
{Armstrong} B.,  {For} B.~Q.,   {Bekki} K.,  2018, \mn@doi [\mnras]
  {10.1093/mnras/sty2445}, \href
  {https://ui.adsabs.harvard.edu/abs/2018MNRAS.481.3651A} {481, 3651}

\bibitem[\protect\citeauthoryear{{Barth}, {Gerber}, {Boberg}, {Friel}  \&
  {Vesperini}}{{Barth} et~al.}{2020}]{K21_Barth_et_al2020}
{Barth} N.~A.,  {Gerber} J.~M.,  {Boberg} O.~M.,  {Friel} E.~D.,   {Vesperini}
  E.,  2020, \mn@doi [\mnras] {10.1093/mnras/staa1019}, \href
  {https://ui.adsabs.harvard.edu/abs/2020MNRAS.494.4548B} {494, 4548}

\bibitem[\protect\citeauthoryear{{Baumgardt}, {Parmentier}, {Anders}  \&
  {Grebel}}{{Baumgardt} et~al.}{2013}]{K42_Baumgardt_et_al2013}
{Baumgardt} H.,  {Parmentier} G.,  {Anders} P.,   {Grebel} E.~K.,  2013,
  \mn@doi [\mnras] {10.1093/mnras/sts667}, \href
  {https://ui.adsabs.harvard.edu/abs/2013MNRAS.430..676B} {430, 676}

\bibitem[\protect\citeauthoryear{{Bekki}}{{Bekki}}{2010}]{K26_Bekki2010}
{Bekki} K.,  2010, \mn@doi [\apjl] {10.1088/2041-8205/724/1/L99}, \href
  {https://ui.adsabs.harvard.edu/abs/2010ApJ...724L..99B} {724, L99}

\bibitem[\protect\citeauthoryear{Bekki}{Bekki}{2013}]{K5_Bekki2013}
Bekki K.,  2013, \mn@doi [Monthly Notices of the Royal Astronomical Society]
  {10.1093/mnras/stt589}, 432, 2298–2323

\bibitem[\protect\citeauthoryear{Bekki}{Bekki}{2015}]{K6_Bekki2015}
Bekki K.,  2015, \mn@doi [Monthly Notices of the Royal Astronomical Society]
  {10.1093/mnras/stv165}, 449, 1625–1649

\bibitem[\protect\citeauthoryear{{Bekki}}{{Bekki}}{2019}]{K27_Bekki2019}
{Bekki} K.,  2019, \mn@doi [\aap] {10.1051/0004-6361/201629898}, \href
  {https://ui.adsabs.harvard.edu/abs/2019A&A...622A..53B} {622, A53}

\bibitem[\protect\citeauthoryear{{Bekki} \& {Chiba}}{{Bekki} \&
  {Chiba}}{2005}]{M8_Bekki_Chiba2005}
{Bekki} K.,  {Chiba} M.,  2005, \mn@doi [\mnras]
  {10.1111/j.1365-2966.2004.08510.x}, \href
  {https://ui.adsabs.harvard.edu/abs/2005MNRAS.356..680B} {356, 680}

\bibitem[\protect\citeauthoryear{{Bekki} \& {Chiba}}{{Bekki} \&
  {Chiba}}{2007a}]{29Bekki_Chiba2007}
{Bekki} K.,  {Chiba} M.,  2007a, \mn@doi [\pasa] {10.1071/AS06023}, \href
  {https://ui.adsabs.harvard.edu/abs/2007PASA...24...21B} {24, 21}

\bibitem[\protect\citeauthoryear{{Bekki} \& {Chiba}}{{Bekki} \&
  {Chiba}}{2007b}]{K35_Bekki_Chiba2007}
{Bekki} K.,  {Chiba} M.,  2007b, \mn@doi [\mnras]
  {10.1111/j.1745-3933.2007.00357.x}, \href
  {https://ui.adsabs.harvard.edu/abs/2007MNRAS.381L..16B} {381, L16}

\bibitem[\protect\citeauthoryear{Bekki \& Chiba}{Bekki \&
  Chiba}{2008}]{K2_Bekki_Chiba2008}
Bekki K.,  Chiba M.,  2008, \mn@doi [The Astrophysical Journal]
  {10.1086/589441}, 679, L89–L92

\bibitem[\protect\citeauthoryear{{Bekki} \& {Chiba}}{{Bekki} \&
  {Chiba}}{2009}]{17Bekki_Chiba2009}
{Bekki} K.,  {Chiba} M.,  2009, \mn@doi [\pasa] {10.1071/AS08020}, \href
  {https://ui.adsabs.harvard.edu/abs/2009PASA...26...48B} {26, 48}

\bibitem[\protect\citeauthoryear{{Bekki} \& {Mackey}}{{Bekki} \&
  {Mackey}}{2009}]{K32_Bekki_Mackey2009}
{Bekki} K.,  {Mackey} A.~D.,  2009, \mn@doi [\mnras]
  {10.1111/j.1365-2966.2008.14320.x}, \href
  {https://ui.adsabs.harvard.edu/abs/2009MNRAS.394..124B} {394, 124}

\bibitem[\protect\citeauthoryear{Bekki \& Stanimirović}{Bekki \&
  Stanimirović}{2009}]{K1_Bekki_Stanimirovic2009}
Bekki K.,  Stanimirović S.,  2009, \mn@doi [Monthly Notices of the Royal
  Astronomical Society] {10.1111/j.1365-2966.2009.14514.x}, 395, 342

\bibitem[\protect\citeauthoryear{{Bekki}, {Beasley}, {Forbes}  \&
  {Couch}}{{Bekki} et~al.}{2004}]{79Bekki_et_al2004}
{Bekki} K.,  {Beasley} M.~A.,  {Forbes} D.~A.,   {Couch} W.~J.,  2004, \mn@doi
  [\apj] {10.1086/381171}, \href
  {https://ui.adsabs.harvard.edu/abs/2004ApJ...602..730B} {602, 730}

\bibitem[\protect\citeauthoryear{{Bergvall}, {Laurikainen}  \&
  {Aalto}}{{Bergvall} et~al.}{2003}]{21Bergvall_Laurikainen_Aalto2003}
{Bergvall} N.,  {Laurikainen} E.,   {Aalto} S.,  2003, \mn@doi [\aap]
  {10.1051/0004-6361:20030542}, \href
  {https://ui.adsabs.harvard.edu/abs/2003A&A...405...31B} {405, 31}

\bibitem[\protect\citeauthoryear{{Besla}, {Kallivayalil}, {Hernquist}, {van der
  Marel}, {Cox}  \& {Kere{\v{s}}}}{{Besla} et~al.}{2012}]{K13_Besla_et_al2012}
{Besla} G.,  {Kallivayalil} N.,  {Hernquist} L.,  {van der Marel} R.~P.,  {Cox}
  T.~J.,   {Kere{\v{s}}} D.,  2012, \mn@doi [\mnras]
  {10.1111/j.1365-2966.2012.20466.x}, \href
  {https://ui.adsabs.harvard.edu/abs/2012MNRAS.421.2109B} {421, 2109}

\bibitem[\protect\citeauthoryear{Bianchini, van der Marel, del Pino, Watkins,
  Bellini, Fardal, Libralato  \& Sills}{Bianchini
  et~al.}{2018}]{K20_Bianchini_et_al2018}
Bianchini P.,  van der Marel R.~P.,  del Pino A.,  Watkins L.~L.,  Bellini A.,
   Fardal M.~A.,  Libralato M.,   Sills A.,  2018, \mn@doi [Monthly Notices of
  the Royal Astronomical Society] {10.1093/mnras/sty2365}, 481, 2125–2139

\bibitem[\protect\citeauthoryear{{Bica}, {Westera}, {Kerber}, {Dias}, {Maia},
  {Santos}, {Barbuy}  \& {Oliveira}}{{Bica} et~al.}{2020}]{K41_Bica_et_al2020}
{Bica} E.,  {Westera} P.,  {Kerber} L. d.~O.,  {Dias} B.,  {Maia} F.,  {Santos}
  Jo{\~a}o F.~C. J.,  {Barbuy} B.,   {Oliveira} R. A.~P.,  2020, \mn@doi [\aj]
  {10.3847/1538-3881/ab6595}, \href
  {https://ui.adsabs.harvard.edu/abs/2020AJ....159...82B} {159, 82}

\bibitem[\protect\citeauthoryear{Bitsakis, Gonz{\'a}lez-L{\'o}pezlira, Bonfini,
  Bruzual, Maravelias, Zaritsky, Charlot  \& Ram{\'i}rez-Siordia}{Bitsakis
  et~al.}{2018}]{11Bitsakis_et_al2018}
Bitsakis T.,  Gonz{\'a}lez-L{\'o}pezlira R.~A.,  Bonfini P.,  Bruzual G.,
  Maravelias G.,  Zaritsky D.,  Charlot S.,   Ram{\'i}rez-Siordia V.~H.,  2018,
  \mn@doi [{The Astrophysical Journal}] {10.3847/1538-4357/aaa244}, 853

\bibitem[\protect\citeauthoryear{{Boutloukos} \& {Lamers}}{{Boutloukos} \&
  {Lamers}}{2003}]{4Boutloukos_Lamers2003}
{Boutloukos} S.~G.,  {Lamers} H.~J.~G.~L.~M.,  2003, \mn@doi [\mnras]
  {10.1046/j.1365-8711.2003.06083.x}, \href
  {https://ui.adsabs.harvard.edu/abs/2003MNRAS.338..717B} {338, 717}

\bibitem[\protect\citeauthoryear{{Chiosi}, {Vallenari}, {Held}, {Rizzi}  \&
  {Moretti}}{{Chiosi} et~al.}{2006}]{16Chiosi_et_al2006}
{Chiosi} E.,  {Vallenari} A.,  {Held} E.~V.,  {Rizzi} L.,   {Moretti} A.,
  2006, \mn@doi [\aap] {10.1051/0004-6361:20054559}, \href
  {https://ui.adsabs.harvard.edu/abs/2006A&A...452..179C} {452, 179}

\bibitem[\protect\citeauthoryear{{Choudhury}, {Subramaniam}, {Cole}  \&
  {Sohn}}{{Choudhury} et~al.}{2018}]{K22_Choudhury_et_al2018}
{Choudhury} S.,  {Subramaniam} A.,  {Cole} A.~A.,   {Sohn} Y.~J.,  2018,
  \mn@doi [\mnras] {10.1093/mnras/sty087}, \href
  {https://ui.adsabs.harvard.edu/abs/2018MNRAS.475.4279C} {475, 4279}

\bibitem[\protect\citeauthoryear{{Cullinane}, {Mackey}, {Da Costa}, {Erkal},
  {Koposov}  \& {Belokurov}}{{Cullinane} et~al.}{2021}]{M6_Cullinane_et_al2021}
{Cullinane} L.~R.,  {Mackey} A.~D.,  {Da Costa} G.~S.,  {Erkal} D.,  {Koposov}
  S.~E.,   {Belokurov} V.,  2021, arXiv e-prints, \href
  {https://ui.adsabs.harvard.edu/abs/2021arXiv210603274C} {p. arXiv:2106.03274}

\bibitem[\protect\citeauthoryear{{D'Onghia} \& {Fox}}{{D'Onghia} \&
  {Fox}}{2016}]{K43_DOnghia_Fox2016}
{D'Onghia} E.,  {Fox} A.~J.,  2016, \mn@doi [\araa]
  {10.1146/annurev-astro-081915-023251}, \href
  {https://ui.adsabs.harvard.edu/abs/2016ARA&A..54..363D} {54, 363}

\bibitem[\protect\citeauthoryear{{Da Costa}}{{Da
  Costa}}{1991}]{K14_Da_Costa1991}
{Da Costa} G.~S.,  1991, in {Haynes} R.,  {Milne} D.,  eds, ~ Vol. 148, The
  Magellanic Clouds. p.~183

\bibitem[\protect\citeauthoryear{{Da Costa}}{{Da Costa}}{2002}]{5Da_Costa2002}
{Da Costa} G.~S.,  2002, in {Geisler} D.~P.,  {Grebel} E.~K.,   {Minniti} D.,
  eds, ~ Vol. 207, Extragalactic Star Clusters. p.~83 (\mn@eprint {arXiv}
  {astro-ph/0106122})

\bibitem[\protect\citeauthoryear{{Dalessandro}, {Lapenna}, {Mucciarelli},
  {Origlia}, {Ferraro}  \& {Lanzoni}}{{Dalessandro}
  et~al.}{2016}]{M5_Dalessandro_et_al2016}
{Dalessandro} E.,  {Lapenna} E.,  {Mucciarelli} A.,  {Origlia} L.,  {Ferraro}
  F.~R.,   {Lanzoni} B.,  2016, \mn@doi [\apj] {10.3847/0004-637X/829/2/77},
  \href {https://ui.adsabs.harvard.edu/abs/2016ApJ...829...77D} {829, 77}

\bibitem[\protect\citeauthoryear{{Dias}, {Kerber}, {Barbuy}, {Bica}  \&
  {Ortolani}}{{Dias} et~al.}{2016}]{K40_Dias_et_al2016}
{Dias} B.,  {Kerber} L.,  {Barbuy} B.,  {Bica} E.,   {Ortolani} S.,  2016,
  \mn@doi [\aap] {10.1051/0004-6361/201527558}, \href
  {https://ui.adsabs.harvard.edu/abs/2016A&A...591A..11D} {591, A11}

\bibitem[\protect\citeauthoryear{Diaz \& Bekki}{Diaz \&
  Bekki}{2012}]{K3_Diaz_Bekki2012}
Diaz J.~D.,  Bekki K.,  2012, \mn@doi [The Astrophysical Journal]
  {10.1088/0004-637x/750/1/36}, 750, 36

\bibitem[\protect\citeauthoryear{{Elson}, {Fall}  \& {Freeman}}{{Elson}
  et~al.}{1987}]{K31Elson_Fall_Freeman1987}
{Elson} R. A.~W.,  {Fall} S.~M.,   {Freeman} K.~C.,  1987, \mn@doi [\apj]
  {10.1086/165807}, \href
  {https://ui.adsabs.harvard.edu/abs/1987ApJ...323...54E} {323, 54}

\bibitem[\protect\citeauthoryear{{Erkal} \& {Belokurov}}{{Erkal} \&
  {Belokurov}}{2020}]{R2Erkal_Belokurov2020}
{Erkal} D.,  {Belokurov} V.~A.,  2020, \mn@doi [\mnras]
  {10.1093/mnras/staa1238}, \href
  {https://ui.adsabs.harvard.edu/abs/2020MNRAS.495.2554E} {495, 2554}

\bibitem[\protect\citeauthoryear{{Erkal} et~al.,}{{Erkal}
  et~al.}{2019}]{R1Erkal_et_al2019}
{Erkal} D.,  et~al., 2019, \mn@doi [\mnras] {10.1093/mnras/stz1371}, \href
  {https://ui.adsabs.harvard.edu/abs/2019MNRAS.487.2685E} {487, 2685}

\bibitem[\protect\citeauthoryear{{Gaia Collaboration} et~al.,}{{Gaia
  Collaboration} et~al.}{2021}]{GaiaCollaboration_2021}
{Gaia Collaboration} et~al., 2021, \mn@doi [\aap]
  {10.1051/0004-6361/202039588}, \href
  {https://ui.adsabs.harvard.edu/abs/2021A&A...649A...7G} {649, A7}

\bibitem[\protect\citeauthoryear{{Gardiner} \& {Noguchi}}{{Gardiner} \&
  {Noguchi}}{1996}]{K15_Gardiner_Noguchi1996}
{Gardiner} L.~T.,  {Noguchi} M.,  1996, \mn@doi [\mnras]
  {10.1093/mnras/278.1.191}, \href
  {https://ui.adsabs.harvard.edu/abs/1996MNRAS.278..191G} {278, 191}

\bibitem[\protect\citeauthoryear{{Glatt} et~al.,}{{Glatt}
  et~al.}{2008}]{M3_Glatt_et_al2008}
{Glatt} K.,  et~al., 2008, \mn@doi [\aj] {10.1088/0004-6256/135/4/1106}, \href
  {https://ui.adsabs.harvard.edu/abs/2008AJ....135.1106G} {135, 1106}

\bibitem[\protect\citeauthoryear{{Glatt}, {Grebel}  \& {Koch}}{{Glatt}
  et~al.}{2010}]{40Glatt_Grebel_Koch2010}
{Glatt} K.,  {Grebel} E.~K.,   {Koch} A.,  2010, \mn@doi [\aap]
  {10.1051/0004-6361/201014187}, \href
  {https://ui.adsabs.harvard.edu/abs/2010A&A...517A..50G} {517, A50}

\bibitem[\protect\citeauthoryear{{Grebel}, {Zaritsky}, {Harris}  \&
  {Thompson}}{{Grebel} et~al.}{1999}]{27Grebel_et_al1999}
{Grebel} E.~K.,  {Zaritsky} D.,  {Harris} J.,   {Thompson} I.,  1999, in {Chu}
  Y.~H.,  {Suntzeff} N.,  {Hesser} J.,   {Bohlender} D.,  eds, ~ Vol. 190, New
  Views of the Magellanic Clouds. p.~405

\bibitem[\protect\citeauthoryear{Harris \& Zaritsky}{Harris \&
  Zaritsky}{2004}]{20Harris_Zaritsky2004}
Harris J.,  Zaritsky D.,  2004, \mn@doi [The Astronomical Journal]
  {10.1086/381953}, 127, 1531–1544

\bibitem[\protect\citeauthoryear{{Hill} \& {Zaritsky}}{{Hill} \&
  {Zaritsky}}{2006}]{85Hill_Zaritsky2006}
{Hill} A.,  {Zaritsky} D.,  2006, \mn@doi [\aj] {10.1086/498647}, \href
  {https://ui.adsabs.harvard.edu/abs/2006AJ....131..414H} {131, 414}

\bibitem[\protect\citeauthoryear{Hunter}{Hunter}{2007}]{matplotlibHunter:2007}
Hunter J.~D.,  2007, \mn@doi [Computing in Science \& Engineering]
  {10.1109/MCSE.2007.55}, 9, 90

\bibitem[\protect\citeauthoryear{{Kamann} et~al.,}{{Kamann}
  et~al.}{2020}]{K24_Kamann_et_al2020}
{Kamann} S.,  et~al., 2020, \mn@doi [\mnras] {10.1093/mnras/stz3506}, \href
  {https://ui.adsabs.harvard.edu/abs/2020MNRAS.492..966K} {492, 966}

\bibitem[\protect\citeauthoryear{{Keller}, {Mackey}  \& {Da Costa}}{{Keller}
  et~al.}{2011}]{K19_Keller_Mackey_Da_Costa2011}
{Keller} S.~C.,  {Mackey} A.~D.,   {Da Costa} G.~S.,  2011, \mn@doi [\apj]
  {10.1088/0004-637X/731/1/22}, \href
  {https://ui.adsabs.harvard.edu/abs/2011ApJ...731...22K} {731, 22}

\bibitem[\protect\citeauthoryear{{Kennicutt}}{{Kennicutt}}{1998}]{K7_Kennicutt1998}
{Kennicutt} Robert~C. J.,  1998, \mn@doi [\apj] {10.1086/305588}, \href
  {https://ui.adsabs.harvard.edu/abs/1998ApJ...498..541K} {498, 541}

\bibitem[\protect\citeauthoryear{{Knapen} \& {Querejeta}}{{Knapen} \&
  {Querejeta}}{2015}]{23Knapen_Querejeta2015}
{Knapen} J.,  {Querejeta} M.,  2015, \mn@doi [Galaxies]
  {10.3390/galaxies3040220}, \href
  {https://ui.adsabs.harvard.edu/abs/2015Galax...3..220K} {3, 220}

\bibitem[\protect\citeauthoryear{{Kumai}, {Basu}  \& {Fujimoto}}{{Kumai}
  et~al.}{1993}]{73Kumai_Basu_Fujimoto1993}
{Kumai} Y.,  {Basu} B.,   {Fujimoto} M.,  1993, \mn@doi [\apj]
  {10.1086/172265}, \href
  {https://ui.adsabs.harvard.edu/abs/1993ApJ...404..144K} {404, 144}

\bibitem[\protect\citeauthoryear{{Kuzma}, {Da Costa}  \& {Mackey}}{{Kuzma}
  et~al.}{2018}]{M1_Kuzma_Da_costa_Mackey2018}
{Kuzma} P.~B.,  {Da Costa} G.~S.,   {Mackey} A.~D.,  2018, \mn@doi [\mnras]
  {10.1093/mnras/stx2353}, \href
  {https://ui.adsabs.harvard.edu/abs/2018MNRAS.473.2881K} {473, 2881}

\bibitem[\protect\citeauthoryear{{Lamers}, {Gieles}  \& {Portegies
  Zwart}}{{Lamers} et~al.}{2005}]{K39_Lamers_Gieles_Portegies_Zwart2005}
{Lamers} H.~J.~G.~L.~M.,  {Gieles} M.,   {Portegies Zwart} S.~F.,  2005,
  \mn@doi [\aap] {10.1051/0004-6361:20041476}, \href
  {https://ui.adsabs.harvard.edu/abs/2005A&A...429..173L} {429, 173}

\bibitem[\protect\citeauthoryear{{Mackey} \& {Gilmore}}{{Mackey} \&
  {Gilmore}}{2003}]{M2_Mackey_Gilmore2003}
{Mackey} A.~D.,  {Gilmore} G.~F.,  2003, \mn@doi [\mnras]
  {10.1046/j.1365-8711.2003.06022.x}, \href
  {https://ui.adsabs.harvard.edu/abs/2003MNRAS.338..120M} {338, 120}

\bibitem[\protect\citeauthoryear{{Mackey}, {Da Costa}, {Ferguson}  \&
  {Yong}}{{Mackey} et~al.}{2013}]{K18_Mackey_et_al2013}
{Mackey} A.~D.,  {Da Costa} G.~S.,  {Ferguson} A.~M.~N.,   {Yong} D.,  2013,
  \mn@doi [\apj] {10.1088/0004-637X/762/1/65}, \href
  {https://ui.adsabs.harvard.edu/abs/2013ApJ...762...65M} {762, 65}

\bibitem[\protect\citeauthoryear{Maji, Zhu, Li, Charlton, Hernquist  \&
  Knebe}{Maji et~al.}{2017}]{37Maji_et_al2017}
Maji M.,  Zhu Q.,  Li Y.,  Charlton J.,  Hernquist L.,   Knebe A.,  2017,
  \mn@doi [The Astrophysical Journal] {10.3847/1538-4357/aa7aa1}, 844, 108

\bibitem[\protect\citeauthoryear{{Mapelli}}{{Mapelli}}{2017}]{K28_Mapelli2017}
{Mapelli} M.,  2017, \mn@doi [\mnras] {10.1093/mnras/stx304}, \href
  {https://ui.adsabs.harvard.edu/abs/2017MNRAS.467.3255M} {467, 3255}

\bibitem[\protect\citeauthoryear{{Mastrobuono-Battisti} \&
  {Perets}}{{Mastrobuono-Battisti} \&
  {Perets}}{2016}]{K29_Mastrobuono_Battisti_Perets2016}
{Mastrobuono-Battisti} A.,  {Perets} H.~B.,  2016, \mn@doi [\apj]
  {10.3847/0004-637X/823/1/61}, \href
  {https://ui.adsabs.harvard.edu/abs/2016ApJ...823...61M} {823, 61}

\bibitem[\protect\citeauthoryear{{Mastropietro}, {Burkert}  \&
  {Moore}}{{Mastropietro} et~al.}{2009}]{K16_Mastropietro_Burkert_Moore2009}
{Mastropietro} C.,  {Burkert} A.,   {Moore} B.,  2009, \mn@doi [\mnras]
  {10.1111/j.1365-2966.2009.15406.x}, \href
  {https://ui.adsabs.harvard.edu/abs/2009MNRAS.399.2004M} {399, 2004}

\bibitem[\protect\citeauthoryear{{McKenzie} \& {Bekki}}{{McKenzie} \&
  {Bekki}}{2018}]{K34_McKenzie_Bekki2018}
{McKenzie} M.,  {Bekki} K.,  2018, \mn@doi [\mnras] {10.1093/mnras/sty1557},
  \href {https://ui.adsabs.harvard.edu/abs/2018MNRAS.479.3126M} {479, 3126}

\bibitem[\protect\citeauthoryear{{McKenzie} \& {Bekki}}{{McKenzie} \&
  {Bekki}}{2021a}]{K23_McKenzie_Bekki2021}
{McKenzie} M.,  {Bekki} K.,  2021a, \mn@doi [\mnras] {10.1093/mnras/staa3376},
  \href {https://ui.adsabs.harvard.edu/abs/2021MNRAS.500.4578M} {500, 4578}

\bibitem[\protect\citeauthoryear{{McKenzie} \& {Bekki}}{{McKenzie} \&
  {Bekki}}{2021b}]{M7_McKenzie_Bekki_2021}
{McKenzie} M.,  {Bekki} K.,  2021b, \mn@doi [\mnras] {10.1093/mnras/stab2171},
  \href {https://ui.adsabs.harvard.edu/abs/2021MNRAS.507..834M} {507, 834}

\bibitem[\protect\citeauthoryear{{Miyamoto} \& {Nagai}}{{Miyamoto} \&
  {Nagai}}{1975}]{K17_Miyamoto_Nagai1975}
{Miyamoto} M.,  {Nagai} R.,  1975, \pasj, \href
  {https://ui.adsabs.harvard.edu/abs/1975PASJ...27..533M} {27, 533}

\bibitem[\protect\citeauthoryear{{Murai} \& {Fujimoto}}{{Murai} \&
  {Fujimoto}}{1980}]{78Murai_Fujimoto1980}
{Murai} T.,  {Fujimoto} M.,  1980, \pasj, \href
  {https://ui.adsabs.harvard.edu/abs/1980PASJ...32..581M} {32, 581}

\bibitem[\protect\citeauthoryear{{Navarro}, {Frenk}  \& {White}}{{Navarro}
  et~al.}{1996}]{K4_Navarro_Frenk_White1996}
{Navarro} J.~F.,  {Frenk} C.~S.,   {White} S. D.~M.,  1996, \mn@doi [\apj]
  {10.1086/177173}, \href
  {https://ui.adsabs.harvard.edu/abs/1996ApJ...462..563N} {462, 563}

\bibitem[\protect\citeauthoryear{{Nayak}, {Subramaniam}, {Choudhury}  \&
  {Sagar}}{{Nayak} et~al.}{2018}]{39Nayak_et_al2018}
{Nayak} P.~K.,  {Subramaniam} A.,  {Choudhury} S.,   {Sagar} R.,  2018, \mn@doi
  [\aap] {10.1051/0004-6361/201732227}, \href
  {https://ui.adsabs.harvard.edu/abs/2018A&A...616A.187N} {616, A187}

\bibitem[\protect\citeauthoryear{{Olsen}, {Zaritsky}, {Blum}, {Boyer}  \&
  {Gordon}}{{Olsen} et~al.}{2011}]{18Olsen_et_al2011}
{Olsen} K. A.~G.,  {Zaritsky} D.,  {Blum} R.~D.,  {Boyer} M.~L.,   {Gordon}
  K.~D.,  2011, \mn@doi [\apj] {10.1088/0004-637X/737/1/29}, \href
  {https://ui.adsabs.harvard.edu/abs/2011ApJ...737...29O} {737, 29}

\bibitem[\protect\citeauthoryear{Osman, Bekki  \& Cortese}{Osman
  et~al.}{2020}]{K10_Osman_Bekki_Cortese2020}
Osman O.,  Bekki K.,   Cortese L.,  2020, \mn@doi [Monthly Notices of the Royal
  Astronomical Society] {10.1093/mnras/staa1554}, 497, 2002–2017

\bibitem[\protect\citeauthoryear{{Pardy}, {D'Onghia}  \& {Fox}}{{Pardy}
  et~al.}{2018}]{89Pardy_DOnghia_Fox2018}
{Pardy} S.~A.,  {D'Onghia} E.,   {Fox} A.~J.,  2018, \mn@doi [\apj]
  {10.3847/1538-4357/aab95b}, \href
  {https://ui.adsabs.harvard.edu/abs/2018ApJ...857..101P} {857, 101}

\bibitem[\protect\citeauthoryear{{Parisi} et~al.,}{{Parisi}
  et~al.}{2015}]{K37_Parisi_et_al2015}
{Parisi} M.~C.,  et~al., 2015, in {Points} S.,  {Kunder} A.,  eds,
  Astronomical Society of the Pacific Conference Series Vol. 491, Fifty Years
  of Wide Field Studies in the Southern Hemisphere: Resolved Stellar
  Populations of the Galactic Bulge and Magellanic Clouds. p.~241

\bibitem[\protect\citeauthoryear{{Parmentier}, {de Grijs}  \&
  {Gilmore}}{{Parmentier} et~al.}{2003}]{3Parmentier_de_Grijs_Gilmore2003}
{Parmentier} G.,  {de Grijs} R.,   {Gilmore} G.,  2003, \mn@doi [\mnras]
  {10.1046/j.1365-8711.2003.06530.x}, \href
  {https://ui.adsabs.harvard.edu/abs/2003MNRAS.342..208P} {342, 208}

\bibitem[\protect\citeauthoryear{{Patel} et~al.,}{{Patel}
  et~al.}{2020}]{M9_Patel_et_al2020}
{Patel} E.,  et~al., 2020, \mn@doi [\apj] {10.3847/1538-4357/ab7b75}, \href
  {https://ui.adsabs.harvard.edu/abs/2020ApJ...893..121P} {893, 121}

\bibitem[\protect\citeauthoryear{{Pe{\~n}arrubia}, {G{\'o}mez}, {Besla},
  {Erkal}  \& {Ma}}{{Pe{\~n}arrubia} et~al.}{2016}]{R3Pernarrubia_et_al2016}
{Pe{\~n}arrubia} J.,  {G{\'o}mez} F.~A.,  {Besla} G.,  {Erkal} D.,   {Ma}
  Y.-Z.,  2016, \mn@doi [\mnras] {10.1093/mnrasl/slv160}, \href
  {https://ui.adsabs.harvard.edu/abs/2016MNRAS.456L..54P} {456, L54}

\bibitem[\protect\citeauthoryear{{Perren}, {Piatti}  \& {V{\'a}zquez}}{{Perren}
  et~al.}{2017}]{56Perren_Piatti_V2017}
{Perren} G.~I.,  {Piatti} A.~E.,   {V{\'a}zquez} R.~A.,  2017, \mn@doi [\aap]
  {10.1051/0004-6361/201629520}, \href
  {https://ui.adsabs.harvard.edu/abs/2017A&A...602A..89P} {602, A89}

\bibitem[\protect\citeauthoryear{{Piatti}}{{Piatti}}{2020}]{53Piatti2020}
{Piatti} A.~E.,  2020, arXiv e-prints, \href
  {https://ui.adsabs.harvard.edu/abs/2020arXiv201212628P} {p. arXiv:2012.12628}

\bibitem[\protect\citeauthoryear{{Piatti}}{{Piatti}}{2021}]{52Piatti2021}
{Piatti} A.~E.,  2021, arXiv e-prints, \href
  {https://ui.adsabs.harvard.edu/abs/2021arXiv210103157P} {p. arXiv:2101.03157}

\bibitem[\protect\citeauthoryear{{Piatti}, {Geisler}, {Sarajedini}, {Gallart}
  \& {Wischnjewsky}}{{Piatti} et~al.}{2008}]{64Piatti_et_al2008}
{Piatti} A.~E.,  {Geisler} D.,  {Sarajedini} A.,  {Gallart} C.,
  {Wischnjewsky} M.,  2008, \mn@doi [\mnras]
  {10.1111/j.1365-2966.2008.13593.x}, \href
  {https://ui.adsabs.harvard.edu/abs/2008MNRAS.389..429P} {389, 429}

\bibitem[\protect\citeauthoryear{{Rafelski} \& {Zaritsky}}{{Rafelski} \&
  {Zaritsky}}{2005}]{1Rafelski_Zaritsky2005}
{Rafelski} M.,  {Zaritsky} D.,  2005, \mn@doi [\aj] {10.1086/424938}, \href
  {https://ui.adsabs.harvard.edu/abs/2005AJ....129.2701R} {129, 2701}

\bibitem[\protect\citeauthoryear{{Santos} Jo{\~a}o F.~C. et~al.,}{{Santos}
  et~al.}{2020}]{K38_Santos_et_al2020}
{Santos} Jo{\~a}o F.~C. J.,  et~al., 2020, \mn@doi [\mnras]
  {10.1093/mnras/staa2425}, \href
  {https://ui.adsabs.harvard.edu/abs/2020MNRAS.498..205S} {498, 205}

\bibitem[\protect\citeauthoryear{{Sawa}, {Fujimoto}  \& {Kumai}}{{Sawa}
  et~al.}{1999}]{67Sawa_Fujimoto_Kumai1999}
{Sawa} T.,  {Fujimoto} M.,   {Kumai} Y.,  1999, in {Chu} Y.~H.,  {Suntzeff} N.,
   {Hesser} J.,   {Bohlender} D.,  eds,  Vol. 190, New Views of the Magellanic
  Clouds. p.~499

\bibitem[\protect\citeauthoryear{{Szigeti}, {M{\'e}sz{\'a}ros}, {Szab{\'o}},
  {Fern{\'a}ndez-Trincado}, {Lane}  \& {Cohen}}{{Szigeti}
  et~al.}{2021}]{K25_Szigeti_et_al2021}
{Szigeti} L.,  {M{\'e}sz{\'a}ros} S.,  {Szab{\'o}} G.~M.,
  {Fern{\'a}ndez-Trincado} J.~G.,  {Lane} R.~R.,   {Cohen} R.~E.,  2021,
  \mn@doi [\mnras] {10.1093/mnras/stab1007}, \href
  {https://ui.adsabs.harvard.edu/abs/2021MNRAS.504.1144S} {504, 1144}

\bibitem[\protect\citeauthoryear{{Tsuge} et~al.,}{{Tsuge}
  et~al.}{2019}]{K36_Tsuge_et_al2019}
{Tsuge} K.,  et~al., 2019, \mn@doi [\apj] {10.3847/1538-4357/aaf4fb}, \href
  {https://ui.adsabs.harvard.edu/abs/2019ApJ...871...44T} {871, 44}

\bibitem[\protect\citeauthoryear{{Tsujimoto}, {Nomoto}, {Yoshii}, {Hashimoto},
  {Yanagida}  \& {Thielemann}}{{Tsujimoto}
  et~al.}{1995}]{K8_Tsujimoto_et_al1995}
{Tsujimoto} T.,  {Nomoto} K.,  {Yoshii} Y.,  {Hashimoto} M.,  {Yanagida} S.,
  {Thielemann} F.~K.,  1995, \mn@doi [\mnras] {10.1093/mnras/277.3.945}, \href
  {https://ui.adsabs.harvard.edu/abs/1995MNRAS.277..945T} {277, 945}

\bibitem[\protect\citeauthoryear{Van Der~Walt, Colbert  \& Varoquaux}{Van
  Der~Walt et~al.}{2011}]{van2011numpy}
Van Der~Walt S.,  Colbert S.~C.,   Varoquaux G.,  2011, Computing in Science \&
  Engineering, 13, 22

\bibitem[\protect\citeauthoryear{{Vesperini}, {Hong}, {Giersz}  \&
  {Hypki}}{{Vesperini} et~al.}{2021}]{K30_Vesperini_et_al2021}
{Vesperini} E.,  {Hong} J.,  {Giersz} M.,   {Hypki} A.,  2021, \mn@doi [\mnras]
  {10.1093/mnras/stab223}, \href
  {https://ui.adsabs.harvard.edu/abs/2021MNRAS.502.4290V} {502, 4290}

\bibitem[\protect\citeauthoryear{{Virtanen} et~al.,}{{Virtanen}
  et~al.}{2020}]{2020SciPy-NMeth}
{Virtanen} P.,  et~al., 2020, \mn@doi [Nature Methods]
  {https://doi.org/10.1038/s41592-019-0686-2}, \href {https://rdcu.be/b08Wh}
  {17, 261}

\bibitem[\protect\citeauthoryear{{Weinberg}}{{Weinberg}}{2000}]{K12_Weinberg2000}
{Weinberg} M.~D.,  2000, \mn@doi [\apj] {10.1086/308600}, \href
  {https://ui.adsabs.harvard.edu/abs/2000ApJ...532..922W} {532, 922}

\bibitem[\protect\citeauthoryear{{Westerlund}}{{Westerlund}}{1997}]{K11_Westerlund1997}
{Westerlund} B.~E.,  1997, {The Magellanic Clouds}

\bibitem[\protect\citeauthoryear{{Yoshizawa} \& {Noguchi}}{{Yoshizawa} \&
  {Noguchi}}{2003}]{69Yoshizawa_Noguchi2003}
{Yoshizawa} A.~M.,  {Noguchi} M.,  2003, \mn@doi [\mnras]
  {10.1046/j.1365-8711.2003.06263.x}, \href
  {https://ui.adsabs.harvard.edu/abs/2003MNRAS.339.1135Y} {339, 1135}

\bibitem[\protect\citeauthoryear{{Zivick} et~al.,}{{Zivick}
  et~al.}{2019}]{24Zivick_et_al2019}
{Zivick} P.,  et~al., 2019, \mn@doi [\apj] {10.3847/1538-4357/ab0554}, \href
  {https://ui.adsabs.harvard.edu/abs/2019ApJ...874...78Z} {874, 78}

\bibitem[\protect\citeauthoryear{{van den Hoek} \& {Groenewegen}}{{van den
  Hoek} \& {Groenewegen}}{1997}]{K9_Van_den_hoek_Groenewegen1997}
{van den Hoek} L.~B.,  {Groenewegen} M.~A.~T.,  1997, \mn@doi [\aaps]
  {10.1051/aas:1997162}, \href
  {https://ui.adsabs.harvard.edu/abs/1997A&AS..123..305V} {123, 305}

\makeatother
\end{thebibliography}




\appendix
\section{SC1 Rotation}

\begin{figure*}
 \includegraphics[width=15cm]{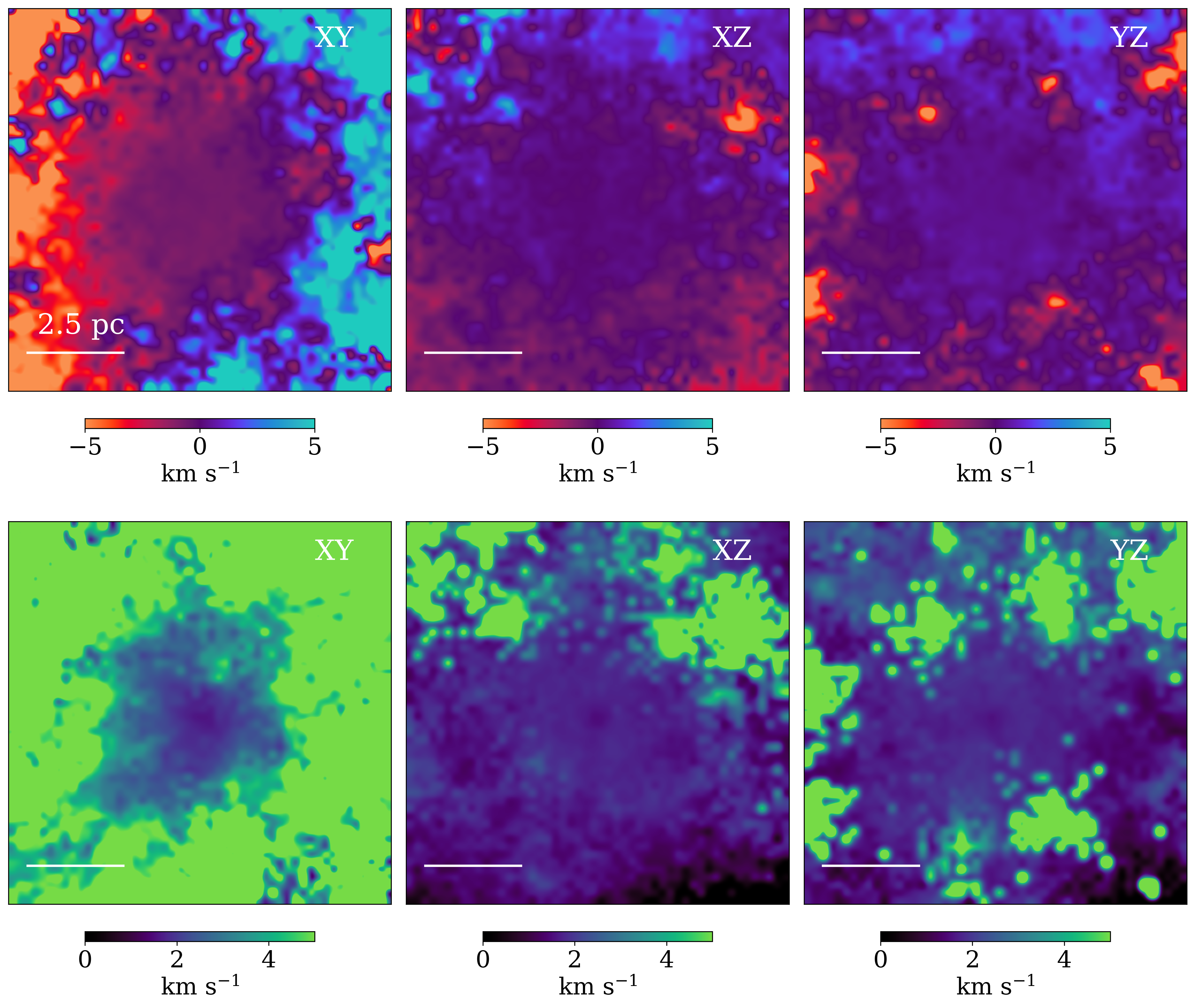}
 \caption{2D maps of line-of-sight velocities centred on the average velocity (upper row) and velocity dispersion (lower row) of new stars projected onto the x-y (left), x-z (middle), and y-z (right) planes for SC1. The velocity has minimally fluctuation within SC1, implying little to no internal rotation.}
 \label{fig:ApRotation}
\end{figure*}
If all stars around a SC are used to derive the 2D maps of velocities for a projection, then the velocity range can be significantly large owing to the steam motions of stars along the line of sight. This large velocity range can make it more difficult to detect possible internal rotation of stars in the SC, in particular, when the SC has a small amount of rotation.  The SC1 does not appear to show global rotation in
the velocity range of -15.9 to 32.4 km s$^{-1}$, but, it could be possible that it has a small amplitude of rotation. In order to investigate, we limit the velocity range to -5 to 5 km s$^{-1}$ and plot the 2D map in Fig \ref{fig:ApRotation}. For clarity, the meshes with V>5 km s$^{-1}$ (V<-5 km s$^{-1}$) are plotted by color for V=5 km s$^{-1}$ (-5 km s$^{-1}$) within a 5 pc radius. Clearly, SC1 has no internal rotation in the three projections, which is consistent with the spherical shape.

\bsp	
\label{lastpage}
\end{document}